# A Large Language Model-Supported Threat Modeling Framework for Transportation Cyber-Physical Systems

**M Sabbir Salek[1], Member, IEEE, Mashrur Chowdhury[1], Senior Member, IEEE, Muhaimin Bin Munir[2], Student Member, IEEE, Yuchen Cai[2], Mohammad Imtiaz Hasan[1], Student Member, IEEE, Jean-Michel Tine[1], Student Member, IEEE, Latifur Khan[2], Fellow, IEEE, and Mizanur Rahman[3], Senior Member, IEEE**

[1]Glenn Department of Civil Engineering, Clemson University, Clemson, SC 29634 USA

[2]Erik Jonsson School of Engineering and Computer Science, The University of Texas at Dallas, Richardson, TX 75080 USA

[3]Department of Civil, Construction, and Environmental Engineering, The University of Alabama, Tuscaloosa, AL 35487 USA

Corresponding author: M Sabbir Salek (e-mail: msalek@clemson.edu).

This work is based upon the work supported by the National Center for Transportation Cybersecurity and Resiliency (TraCR) (a U.S. Department of Transportation National University Transportation Center) headquartered at Clemson University, Clemson, South Carolina, USA. Any opinions, findings, conclusions, and recommendations expressed in this material are those of the author(s) and do not necessarily reflect the views of TraCR, and the U.S. Government assumes no liability for the contents or use thereof.

**ABSTRACT** Increased reliance on automation and connectivity exposes transportation cyber-physical systems (CPS) to many cyber vulnerabilities. Existing threat modeling frameworks are often narrow in scope, labor-intensive, and require substantial cybersecurity expertise. To this end, we introduce the Transportation Cybersecurity and Resiliency Threat Modeling Framework (TraCR-TMF), a large language model (LLM)-based threat modeling framework for transportation CPS that requires limited cybersecurity expert intervention. TraCR-TMF identifies threats, potential attack techniques (i.e., methods to exploit vulnerabilities), and relevant countermeasures (e.g., attack detection and mitigation strategies) for transportation CPS. Three LLM-based approaches support these identifications: (i) a retrieval-augmented generation approach requiring no cybersecurity expert intervention, (ii) an in-context learning approach with low expert intervention, and (iii) a supervised fine-tuning approach with moderate expert intervention. TraCR-TMF offers LLM-based attack path identification for critical assets based on vulnerabilities across transportation CPS entities. Additionally, it incorporates the Common Vulnerability Scoring System (CVSS) scores of known exploited vulnerabilities to prioritize threat mitigations. The framework was evaluated through two cases. First, the framework identified relevant attack techniques for various transportation CPS applications, 73% of which were validated by cybersecurity experts as correct. Second, the framework was used to identify attack paths for a target asset in a real-world cyberattack incident. TraCR-TMF successfully predicted exploitations, like lateral movement of adversaries, data exfiltration, and data encryption for ransomware, as reported in the incident. These findings show TraCR-TMF's efficacy in transportation CPS threat modeling, while reducing the need for extensive involvement of cybersecurity experts. To facilitate real-world adoptions, all our codes are shared via an open-source repository.

## I. INTRODUCTION

With the rise of artificial intelligence (AI), automation, and widespread use of wired and wireless communication technologies, legacy transportation systems are increasingly evolving into transportation cyber-physical systems (CPS). These systems integrate cyber systems, such as onboard, roadside, and cloud-based computing systems, with physical systems like transportation sensors and actuators via network communications. While transportation CPS offer enhanced safety, efficiency, and air quality [1], they also introduce a significantly expanded attack surface for malicious actors [2], [3]. Cyberattacks on these systems can jeopardize human lives and disrupt critical operations.

Recent trends show a sharp rise in cyberattacks targeting multimodal transportation systems. Cyber incidents increased by 400% in road transportation between 2017 and 2022 [4], by 530% in aviation between 2019 and 2020 [5], and by 900% in maritime systems between 2017 and 2021 [6]. These numbers highlight escalating cybersecurity risks. Compounding this concern is the interconnectivity of transportation CPS with other critical infrastructures, such as power grids, electronic banking systems, and healthcare systems, which raises the risk of cascading disruptions from a single cyber incident. Consequently, securing transportation CPS against evolving threats is now a pressing priority for transportation stakeholders, cybersecurity professionals, and policymakers.

Addressing the cybersecurity needs of multimodal transportation CPS requires both proactive and reactive measures. Proactive approaches focus on identifying and mitigating vulnerabilities during system design, while reactive methods handle detection, response, and recovery. Among proactive strategies, threat modeling stands out for systematically identifying threats and vulnerabilities before exploitation [7], [8]. It helps define security requirements, communicate risks, and recommend appropriate cyberattack detection and mitigation strategies. Thus, incorporating threat modeling early in the CPS application development cycle helps reduce risks and improve overall system resilience.

Despite its critical importance, the adoption of threat modeling in the transportation CPS domain remains limited. Although threat modeling has gained significant traction in the software development community, its application to transportation CPS has been relatively underexplored. Existing studies on threat modeling for transportation CPS, which we highlighted in Section III-B of this paper, are often constrained to specific modes of transportation or limited threat scenarios, e.g., [9], [10], [11]. Moreover, traditional frameworks demand a comprehensive understanding of system components, interactions, vulnerabilities, and threats—a process that is time-intensive and complex when done manually. These challenges hinder broader adoption despite established benefits of threat modeling [12].

To address these challenges, this study presents a novel threat modeling framework to help enhance the cybersecurity and resilience of transportation CPS. We present the Transportation Cybersecurity and Resiliency Threat Modeling Framework (TraCR-TMF), which leverages existing cybersecurity tools, established threat models, and the logical reasoning and generative capabilities of large language models (LLMs). TraCR-TMF is a multi-stage threat modeling approach that is specifically tailored to assist cybersecurity professionals dealing with transportation CPS in conducting comprehensive threat assessments and identifying relevant countermeasures that could help detect and mitigate identified threats.

TraCR-TMF offers several key advantages. First, it incorporates three alternative LLM-based approaches to identify relevant attack techniques, i.e., methods to exploit vulnerabilities. These LLM-based approaches can extract relevant details from structured databases, such as the MITRE Adversarial Tactics, Techniques, and Common Knowledge (ATT&CK) matrix [13], with zero to moderate cybersecurity expert interventions. This capability enables the automatic identification of attack techniques and their corresponding detection and mitigation strategies. Second, TraCR-TMF offers autonomous mapping of potential attack paths and associated attack techniques, retrieved from the MITRE ATT&CK matrix. These attack paths, identified by analyzing transportation CPS application vulnerabilities, could lead to the compromise of critical transportation CPS assets. Third, the framework integrates other widely utilized open frameworks, such as the Common Vulnerabilities and Exposures (CVE) [14], the Known Exploited Vulnerabilities (KEV) Catalog [15], and the Common Vulnerability Scoring System (CVSS) [16], to help prioritize threat mitigations based on the severity of associated known exploited vulnerabilities. By integrating these capabilities, TraCR-TMF streamlines the threat modeling process and enhances the accessibility of cybersecurity analysis for transportation CPS.

The rest of this paper is structured as follows: Section II outlines our key contributions. Section III presents a review of related methods and studies. Section IV provides briefs on tools and databases relevant to TraCR-TMF and its evaluation cases considered in this study. Section V details the TraCR-TMF's design and implementation. Section VI presents two evaluation cases that assess the effectiveness of TraCR-TMF in modeling threats to a wide variety of transportation CPS applications. Section VII addresses practical deployment concerns of the framework. Finally, Section VIII concludes the paper by summarizing the findings and the scope for future research.

## II. CONTRIBUTIONS

This study aims to address the need for an easy-to-adapt, cost-effective threat modeling framework for transportation CPS that could assist in identifying vulnerabilities, potential threats, and their available countermeasures. The key contributions of this study are as follows:

- A multi-stage threat modeling framework for transportation CPS, incorporating widely used open-source tools and databases, along with three LLM-supported adoption strategies.
- A retrieval-augmented generation (RAG)-based architecture to identify specific attack techniques that attackers could use to exploit vulnerabilities in a transportation CPS application, along with the corresponding countermeasures, requiring no cybersecurity expert intervention.
- An in-context learning-based approach for attack technique and countermeasure identification, leveraging an engineered prompt and requiring minimal cybersecurity expert intervention.
- A supervised fine-tuning approach to facilitate the identification of relevant attack techniques and corresponding countermeasures, requiring moderate cybersecurity expert intervention.
- An LLM-based, asset-centric threat modeling approach that leverages an engineered prompt to identify potential attack paths and associated attack techniques leading to the compromise of any specified critical asset. This asset-centric threat modeling offers insights into the multi-layered vulnerabilities of transportation CPS, i.e., a chain of vulnerabilities across multiple CPS entities that can be exploited together to perform an attack that aims to compromise one or more critical assets.
- A threat-informed risk mitigation prioritization strategy, considering the severity of underlying known exploited vulnerabilities, i.e., vulnerabilities that have been exploited in previous incidents.

TraCR-TMF enables (i) systematic identification of potential threats, and their categories based on a well-established threat model, known as STRIDE, (ii) identification of known attack techniques, and their existing detection and mitigation strategies from the MITRE ATT&CK matrix based on the STRIDE-identified threats, (iii) identification of potential attack paths and the associated attack techniques for each step of an attack path that could compromise a critical transportation CPS asset, and (iv) prioritization of threat mitigation strategies using open frameworks, such as MITRE ATT&CK, CVE, KEV Catalog, and CVSS.

## III. RELATED WORK

Threat modeling is a well-studied topic in cybersecurity; consequently, many studies focus on different methods of threat modeling. Systematic reviews of these studies have been presented in several survey papers, for example, [17], [18], [19], [20], [21], [22], [23], which categorized the existing threat modeling methods and frameworks from different perspectives, such as reviewing them based on the underlying methods or the type of systems or applications to which threat modeling is applied. In this section, we first present a general overview of different existing threat modeling approaches. Then, we present an overview of the threat modeling studies that particularly relate to transportation CPS and/or leveraged LLMs to automate threat modeling.

### *A. THREAT MODELING IN GENERAL*

Threat modeling or analysis has been defined from different perspectives in literature. Uzunov and Fernandez [24] defined threat modeling as a structured analysis of potential threats or attacks in various contexts, such as analysis of threats or attacks specific to systems and/or technologies. Another study [25] described threat modeling as a technique for identifying and documenting security threats in a software system while systematically uncovering the system's strengths and weaknesses. In [26], Xiong et al. referred to threat modeling as a process for assessing potential threats, risks, and attacks. Although threat and risk are used interchangeably in some studies, there lies a distinction between them. Al-Fedaghi and Alkandari [27] defined risk as a function of vulnerability and threat. Thus, analyzing threats is essential in risk assessment. In contrast, threat modeling is a structured process for identifying, analyzing, and mitigating security threats in a system, application, or network. It helps security professionals understand potential attack vectors, assess risks, and implement appropriate defenses.

The key aspects of threat modeling can include [22], [28]: (i) identifying assets, which implies determining what needs protection, for example, data, hardware, and software; (ii) determining threats, which refers to recognizing potential adversaries and attack vectors; (iii) analyzing potential attack paths, which implies exploring how an attacker could exploit the existing system vulnerabilities; (iv) assessing risks, which involves evaluation of the likelihood and impact of the threats; and (v) identifying and implementing mitigation strategies: this refers to the identification and implementation of appropriate security controls based on different analyzed threats, all of which are captured by the TraCR-TMF.

The existing threat modeling approaches can be broadly categorized as (i) formula-based threat modeling and (ii) model-based threat modeling, as presented in [29]. Formula-based methods are used for threat analysis and risk assessment of a system, primarily utilizing tables, textual descriptions, and mathematical formulas. In contrast, model-based methods are a type of threat analysis approach that employs various models to assess system threats and risks, utilizing data flow diagrams, graphs, and tree models for modeling and analysis.

The formula-based methods can be further classified into (i) asset-centric, (ii) vulnerability-centric, and (iii) attacker-centric threat modeling approaches [29]. An asset-centric approach, which can be thought of as a top-down approach, identifies target assets first and then maps potential attack paths using expert knowledge, which enables early threat mitigations. A popular asset-centric method, OCTAVE, was introduced by Alberts et al. [30] for analyzing threats in a large organization with a multi-layered hierarchy or geographically distributed systems. This method involves three phases, as follows: (i) building asset-centric threat profiles, (ii) identifying infrastructure vulnerabilities, and (iii) developing security strategies and plans. In contrast to an asset-centric approach, a vulnerability-centric threat modeling approach is a bottom-up approach that initiates with a vulnerability of a system and analyzes other potential vulnerabilities or failures that could be caused by that vulnerability. For instance, the Common Vulnerability Scoring System (CVSS) assesses the severity of different system-level vulnerabilities so that appropriate countermeasures can be prioritized accordingly. Finally, an attacker-centric threat modeling approach focuses on analyzing potential attackers by assessing their knowledge, attack paths, motivations, and available resources. This approach enables threat modeling and risk assessment by identifying the root cause of attacks.

The model-based methods can also be further classified into (i) graph-based and (ii) attack tree-based threat modeling approaches [29]. Graph-based approaches utilize graphs with nodes and directional edges, representing quantitative relationships among the nodes. Among the most popular graph-based approaches is STRIDE, a Microsoft-developed threat modeling tool [31], which utilizes six categories of threats, i.e., spoofing, tampering, repudiation, information disclosure, denial-of-service, and elevation of privilege, to conduct threat modeling. LINDDUN, developed by Deng et al. [32], is another popular graph-based threat modeling framework that helps identify privacy-related threats. Another example of a graph-based threat modeling approach is PASTA, which was introduced by UcedaVelez and Morana [33]. PASTA is primarily focused on addressing high-risk threats to optimize investment and resource utilization.

In contrast to graph-based threat modeling, tree-based approaches utilize a tree to represent the hierarchical relationship among nodes. The attack tree is the most used tree-based threat modeling approach. This method represents threats as hierarchical trees, where each branch represents a potential attack path. Attack trees are especially useful in visualizing complex attack scenarios and understanding dependencies. For instance, Saini et al. [34] utilized the attack trees concept, originally developed by Bruce Schneier [35], to conduct threat modeling for a grid computing security subsystem called MyProxy. The MyProxy subsystem, within the Globus grid computing toolkit, serves as an online credential repository and a certificate authority for its users. Saini et al. [34] utilized the SecureITree tool to develop an attack tree for the MyProxy subsystem.

### *B. TRANSPORTATION CPS AND/OR LLM-BASED THREAT MODELING*

Although threat modeling is widely studied for software security in general, its applicability to multimodal transportation systems has been considered in a handful of studies, for example, [9], [11], [36], [37], [38]. In Table I, we present a review of some of these studies, highlighting the threat modeling approaches they presented, along with their specific use case, and their corresponding advantages and limitations. As observed from Table I, the existing threat modeling frameworks for transportation systems are often limited to specific threat types and use cases, and lack automation.

Some recent studies, such as [39], [40], [41], [42], have utilized the advanced reasoning and text generation capabilities of LLMs to automate threat modeling approaches. In Table I, we also present an overview of these studies, along with their limitations, such as applicability to limited use cases, attack scenarios, and high false positive rates, all of which are addressed in this study.

TABLE I
SUMMARY OF RELEVANT EXISTING THREAT MODELING STUDIES

| | Study | Use Case and Threat Modeling Approach | Advantages | Limitations |
|---|---|---|---|---|
| Transportation CPS Threat Modeling | Ramazanzadeh et al. [11] | Security object-oriented colored Petri nets (SOOCPN) for automated threat classification in transportation systems | Automated threat classification into levels and color-coded sublevels based on quantified risk | Relies on user-assigned, arbitrary risk ranges, impacting consistency in real-world use |
| | He et al. [9] | Markov-chain modeling of DNS[a] rebinding attacks in maritime IoT[b] | Effective detection and mitigation of DNS rebinding threats | Focuses solely on DNS rebinding, limiting applicability to other threats |
| | Kumar et al. [36] | LSTM-VAE[c] for feature extraction and Bi-GRU[d] for attack type identification in maritime IoT | Automates manual analysis and addresses low detection and high false-alarm rates | Requires large training datasets and only detects threat types seen during training |
| | Subran et al. [37] | STRIDE-based analysis of EV[e] charging infrastructure with Mininet-based DoS[f] attack simulation | Automates threat modeling of DoS scenarios in a simulated EV network | Limited to DoS threats; other threat categories remain unaddressed |
| | Hamad et al. [38] | Manual integration of subsystem-specific threat models into a unified attack-tree framework for modern vehicles | Produces reusable attack trees applicable across different vehicle systems | Lacks automation; relies entirely on manual literature review and tree generation |
| LLM-based Threat Modeling | Munir et al. [39] | RAG-enhanced, LLM-based cyberattack detection framework for transit security | Leverages MITRE ATT&CK matrix dynamically; no manual ATT&CK lookup required | Evaluation limited to a single transit scenario and no comparison with other LLM-based strategies |
| | Gabrys et al. [40] | LLM-based conversion of IDS[g] rules to human-readable form and mapping to ATT&CK via ChatGPT and BERT | Transforms complex IDS rules into a readable format and links observed behaviors to ATT&CK techniques | Only applicable for IDSs and depends on manually labeled training data |
| | Branescu et al. [41] | BERT-based and ChatGPT-based mapping of CVE descriptions to ATT&CK techniques | Systematic analysis and mapping of vulnerabilities to tactics; assists in prioritizing mitigations | Needs additional data or metadata to improve mapping performance |
| | Nir et al. [42] | ML/DL[h] and behavioral analysis for automated IDS threat labeling | Automates anomaly detection and attack classification | Faces scalability challenges, high false-positive rates, and requires domain-specific adaptation |

[a]DNS: Domain Name System; [b]IoT: Internet-of-Things; [c]LSTM-VAE: Long Short-Term Memory-based Variational Autoencoder; [d]Bi-GRU: Bidirectional Gated Recurrent Unit; [e]EV: Electric Vehicle; [f]DoS: Denial-of-Service; [g]IDS: Intrusion Detection System; [h]ML/DL: Machine Learning or Deep Learning

Based on these reviews, this study is motivated to develop a threat modeling framework for transportation CPS that can be adopted without being constrained by extensive cybersecurity expertise and substantial manual labor requirements. To this end, our TraCR-TMF offers different alternative adoption strategies leveraging different levels of cybersecurity expert intervention and the logical reasoning capabilities of LLMs.

## IV. PRELIMINARIES

In this section, we briefly discuss a few preliminaries that are essential for presenting the TraCR-TMF and the evaluation cases considered in this study. The following subsections provides the readers with quick introductions to (i) the Microsoft Security Development Lifecycle (MS SDL) threat modeling tool, which identifies different types of threats in a transportation CPS, (ii) the MITRE ATT&CK matrix, which serves as a comprehensive cyberattack database, (iii) the national intelligent transportation systems reference architecture, which provides reference architecture, including physical and functional objects, data flow, etc. for various transportation CPS applications, (iv) Commons Vulnerabilities and Exposures (CVE), and (v) Common Vulnerability Scoring System (CVSS).

### A. MS SDL THREAT MODELING TOOL

Developed by Microsoft in 2018, the MS SDL threat modeling tool [31] serves as a core component of MS software development lifecycle, an engineered approach primarily to help web-based application and/or software developers identify cyber vulnerabilities, threats, potential cyberattacks, and their countermeasures. This open-source threat modeling tool allows developers to identify and mitigate potential cybersecurity issues early in their software or application development lifecycle.

MS SDL threat modeling tool utilizes the STRIDE model that categorizes threats into six categories, as follows: (i) spoofing, (ii) tampering, (iii) repudiation, (iv) information disclosure, (v) denial-of-service, and (vi) elevation of privilege [43]. Table II provides a brief description of each of these threat categories. This graph-based threat modeling tool provides users with a graphical user interface (GUI) for developing data flow diagrams (DFDs) using different stencils, such as process, external interactor, data store, data flow, and trust boundary [44]. Once the DFD is provided for an application, the MS SDL threat modeling tool generates a threat report, showcasing the threats associated with each interaction observed in the given DFD based on the STRIDE model. The tool utilizes a "STRIDE per element" approach, which refers to the identification of STRIDE threats for each DFD element. A DFD element can be a process, an external interactor, a data store, or a data flow. For each threat identified by the MS SDL threat modeling tool, the report presents the associated STRIDE threat category, a brief description of the threat, along with its potential mitigations.

### B. MITRE ATT&CK MATRIX

MITRE ATT&CK is a globally recognized cybersecurity database that provides a structured approach to understanding and analyzing cyber threats [13]. Developed by The MITRE Corporation, ATT&CK provides comprehensive documentation of real-world adversary behaviors to help organizations improve their detection, response, and mitigation strategies. This framework is widely used by cybersecurity professionals for threat intelligence, security operations, incident response, and adversary emulation.

MITRE ATT&CK is organized into matrices that cover different environments, including enterprise, mobile, and industrial control systems (ICS). Each matrix outlines various adversary behaviors that attackers use to infiltrate, persist, and execute malicious activities within a network. Each matrix includes some common key components, such as tactics, techniques, sub-techniques, and procedures. Table III presents a brief description of each component, along with some examples. For each technique, the MITRE ATT&CK provides a list of associated sub-techniques and procedure

TABLE II
STRIDE THREAT CATEGORIES (ADOPTED FROM [43])

| Threat Category | Threat Description |
|---|---|
| Spoofing | The unauthorized access and use of another user's authentication credentials, such as a username and password. |
| Tampering | The malicious alteration of data, including unauthorized modifications to persistent data (e.g., in a database) or changes to data in transit over an open network like the Internet. |
| Repudiation | A situation where a user denies performing an action without any means to verify the claim. For example, in a system without proper logging, an unauthorized operation can be executed without accountability. |
| Information disclosure | The unauthorized exposure of information to individuals who should not have access. Examples include users reading files they lack permission for or attackers intercepting data transmitted between systems. |
| Denial of service | The disruption of access to a system or service, rendering it temporarily unavailable or unusable to legitimate users; for example, making a web server inaccessible. |
| Elevation of privilege | An unprivileged user gaining higher-level access, potentially compromising or taking control of the entire system. This includes scenarios where an attacker bypasses all security defenses and integrates into the trusted system, posing a significant threat. |

TABLE III
KEY COMPONENTS OF MITRE ATT&CK (ADOPTED FROM [13])

| Component | Description | Examples |
|---|---|---|
| Tactic | Tactics represent the high-level objectives or goals that an adversary seeks to achieve during an attack. A tactic provides a context on why an attacker is performing a particular action. | • Initial Access – Gaining entry into a system or network (e.g., phishing, exploiting vulnerabilities).<br>• Persistence – Establishing long-term access within a compromised environment. |
| Technique | Techniques describe how an adversary achieves a specific tactic. Each technique represents a method or approach used by attackers to execute their objectives. | • Phishing (T1566) – Sending deceptive emails to trick users into providing credentials.<br>• Process Injection (T1055) – Injecting malicious code into legitimate processes to evade detection. |
| Sub-technique | Sub-techniques break down techniques into more specific methods, offering deeper granularity into how adversaries operate. They help organizations refine their detection and response strategies by addressing precise attack behaviors. | Sub-techniques of phishing (T1566):<br>• Spearphishing Attachment (T1566.001) – Delivering malware via an email attachment.<br>• Spearphishing Link (T1566.002) – Tricking users into clicking a malicious link. |
| Procedure | Procedures define the specific ways in which threat actors or malware implement techniques and sub-techniques. They provide real-world examples of how known adversary groups execute attacks. | Procedures of phishing (T1566):<br>• Axiom (G0001) – Axiom has used spear phishing to initially compromise victims.<br>• Hikit (S0009) – Hikit has been spread through spear phishing |

examples as well as the existing detection and mitigation strategies for those attacks.

### C. NATIONAL REFERENCE ARCHITECTURE FOR TRANSPORTATION CPS APPLICATIONS

Transportation CPS play a pivotal role in modern urban infrastructure by facilitating efficient traffic management, enhancing safety, and improving mobility. These systems rely on a sophisticated network of interconnected physical and functional entities that exchange data to help provide efficient and safe transportation services. To analyze the cybersecurity challenges within transportation CPS, this study utilizes the U.S. Department of Transportation's (USDOT) Architecture Reference for Cooperative and Intelligent Transportation (ARC-IT) [45]. ARC-IT offers a comprehensive framework for transportation CPS design and implementation, encompassing detailed service packages and associated components. Here, we present a brief introduction to the different components of these reference architectures:

#### 1) SERVICE PACKAGES

ARC-IT categorizes transportation CPS functionalities into 156 service packages (to date), distributed across 12 distinct domains such as Public Safety, Data Management, and Vehicle Safety. Each service package on ARC-IT is equivalent to a transportation CPS application that integrates physical systems and functional components in a transportation CPS environment to address specific transportation services. This list of service packages is updated over time to include newly introduced transportation CPS.

#### 2) PHYSICAL AND FUNCTIONAL OBJECTS

Physical objects represent the tangible entities involved in transportation CPS, including centers (e.g., traffic management center, emergency management center, payment administration center), fields (e.g., roadway equipment, intermodal terminal, electric charging station), personal devices (e.g., pedestrian, payment device, remote access device), support systems (e.g., data distribution system, archived data system), and vehicles (e.g., basic transit vehicle, basic commercial vehicle). These objects serve as the core infrastructure of transportation CPS, enabling critical information exchange for operational and security purposes. Each physical object is associated with one or more functional objects, which are deployment-specific units designed to fulfill the functional requirements of the system.

#### 3) INFORMATION AND DATA FLOWS

Information and data flows serve as the backbone of transportation CPS communications, facilitating data exchange among physical and functional objects. Each information flow originates from a physical object, referred to as the initiator or source, and is received by another physical object, referred to as the acceptor or destination. In contrast, data flow refers to the data exchange between two functional objects in a transportation CPS application. According to these definitions of information and data flows used in ARC-IT, each information flow (between two physical objects) can be broken down into several data flows (between two functional objects).

### D. COMMON VULNERABILITIES AND EXPOSURES

Common Vulnerabilities and Exposures (CVE) is a public catalog of known cybersecurity vulnerabilities, providing unique identifiers (CVE IDs) for each listed vulnerability [14]. CVE offers a standardized way to reference vulnerabilities across tools and organizations. For cybersecurity practitioners, CVE is essential for prioritizing risks, coordinating responses, and ensuring consistency in vulnerability tracking. Analysts often use CVE entries to evaluate exploitability, and plan timely threat mitigation strategies. This shared reference system enhances collaboration and accelerates defensive actions across the cybersecurity community.

### E. COMMON VULNERABILITY SCORING SYSTEM

The Common Vulnerability Scoring System (CVSS) is a standardized framework used to assess the severity of software vulnerabilities [16]. Version 3.1 of CVSS includes three metric groups as follows: (i) Base metric group, which reflects the inherent properties of a vulnerability that remain constant over time; (ii) Temporal metric group, which captures factors such as exploit code maturity and remediation level; and (iii) Environmental metric group, which considers the specific context of an organization's information technology (IT) environment [16]. The Base metrics evaluate aspects like how a vulnerability is exploited (attack vector), the complexity of the attack, required privileges, user interaction, and the potential impacts on confidentiality, integrity, and availability. Cybersecurity practitioners can find the CVSS scores for known vulnerabilities in the National Vulnerability Database (NVD), which reports the base score for each CVE ID [46]. These scores help analysts prioritize which vulnerabilities pose the most significant risk based on severity and exploitability. Using CVSS, organizations can make informed decisions in their vulnerability management and threat mitigation efforts.

## V. OVERVIEW OF TRACR-TMF

The TraCR-TMF, presented in Fig. 1, is a multi-stage threat modeling framework that requires limited cybersecurity expert intervention and leverages different LLM-enabled approaches to (i) identify potential attack techniques that could exploit the vulnerabilities within transportation CPS, (ii) analyze these attack techniques to identify potential attack paths leading to critical transportation CPS assets. In addition, the framework helps cybersecurity professionals prioritize threat mitigations by considering the exploitability and the subsequent impact of an exploitation. As depicted in Fig. 1, this multi-stage threat modeling is carried out by augmenting a threat report generated from the MS SDL threat modeling tool with relevant attack techniques and their corresponding detection and mitigation strategies retrieved from the MITRE ATT&CK matrix. Once relevant MITRE ATT&CK techniques have been identified, a customized LLM is utilized to perform an asset-centric threat modeling in which the LLM identifies potential attack paths, along with associated attack techniques, leading to the compromise of specified assets. Relevant MITRE ATT&CK techniques also aid in threat mitigation prioritization by ranking the associated vulnerabilities (as listed in CVE) in descending order of their CVSS scores. Different stages of the TraCR-TMF, presented in Fig. 1, are detailed in the following subsections.

### A. STRIDE-BASED THREAT IDENTIFICATION (STAGE 1)

The TraCR-TMF utilizes the open-source MS SDL threat modeling tool [31] to model threats based on the STRIDE model. MS SDL threat modeling tool requires the systems or reference architecture as a DFD, which is modeled through the tool's GUI. As mentioned in Section IV-A, a DFD can include four types of elements, as follows: (i) Processes: functional objects in a transportation CPS application are modeled as processes; (ii) External interactors: terminators, i.e., entities that are involved (e.g., a human user) but fall outside the design scope of a transportation CPS application, are modeled as external interactors; (iii) Data stores: in-house, external, or cloud-based data stores are modeled as data stores; and (iv) Data flows: interactions between functional objects, or between a functional object and an external interactor, are modeled as data flows. In addition, a trust boundary is used to indicate a physical object involved in a transportation CPS application that may include one or more functional objects.

Once the DFD, along with the trust boundaries, of a transportation CPS application is provided, the MS SDL threat modeling tool generates a STRIDE-based threat report. This report outlines spoofing, tampering, repudiation, information disclosure, denial-of-service, and elevation of privilege-related threats for each interaction in a DFD. These threats are identified for any processes, external interactors, data stores, or data flows associated with each interaction in a DFD. However, this report does not provide information on specific attack techniques associated with each threat that attackers could utilize to exploit the vulnerabilities. Knowledge of these attack techniques, along with their existing detection and mitigation methods, would enable application developers to implement targeted countermeasures for effective prevention, detection, and mitigation of potential threats, which are addressed by stages 2 through 4 of the TraCR-TMF.

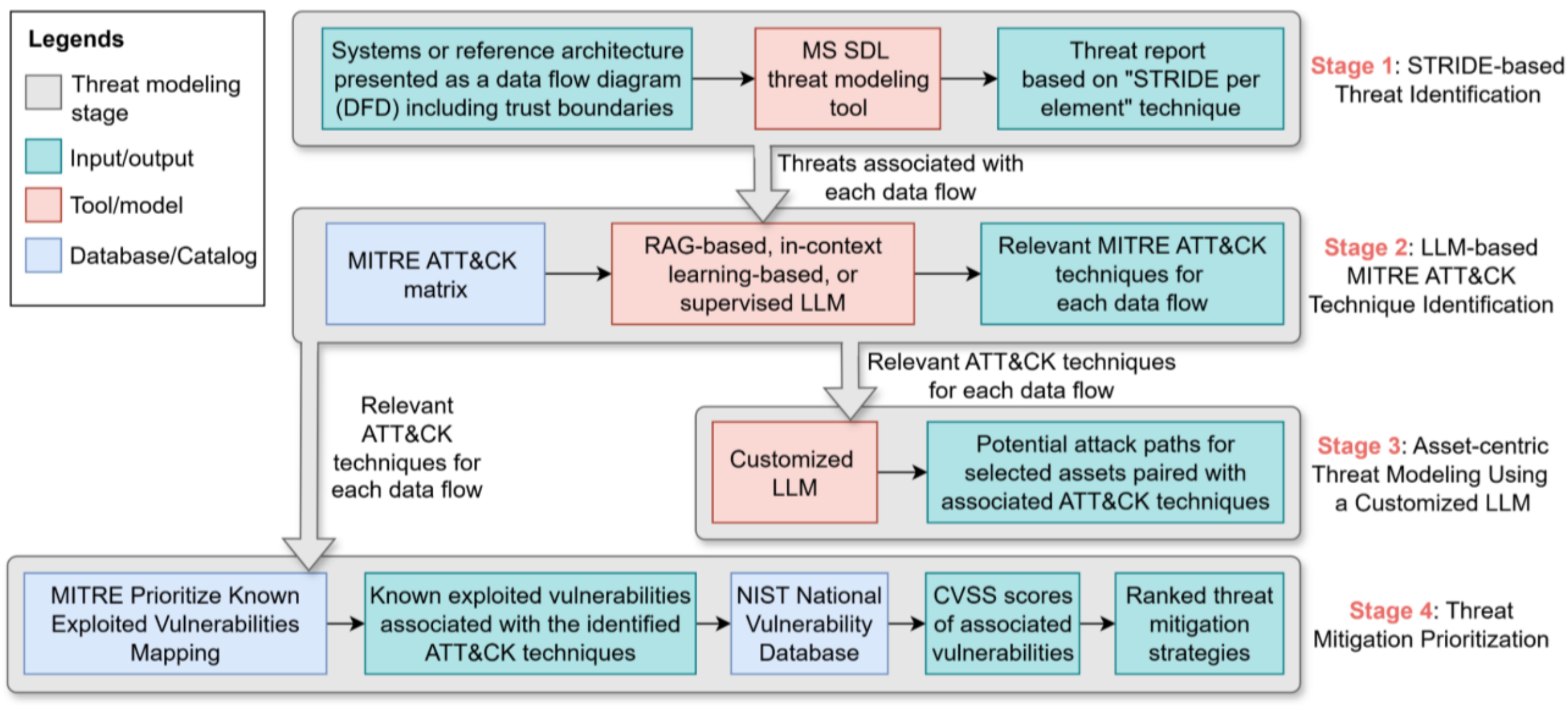

FIGURE 1. Overview of TraCR-TMF for transportation CPS.

### *B. LLM-BASED MITRE ATT&CK TECHNIQUE IDENTIFICATION (STAGE 2)*

The TraCR-TMF utilizes an LLM to identify attack techniques from the MITRE ATT&CK matrix that could be relevant in exploiting vulnerabilities within transportation CPS based on the threats identified by the STRIDE model. In this study, we consider three different approaches with LLMs to map the STRIDE-based threats to the potentially associated MITRE ATT&CK techniques that require different levels of cybersecurity expert intervention. These approaches are as follows: (i) a RAG-based approach with an LLM, (ii) an in-context learning-based approach with an LLM, and (iii) a supervised fine-tuning approach using an LLM. The input that is included in the prompts of all these approaches, which we refer to as the "basic input" in this study, includes the following components (as presented in Fig. 2):

- **Data Flow:** A name is specified for the data flow being analyzed.
- **Data Flow Definition:** An explanation of the data flow is provided
- **Initiator and Acceptor:** Descriptions of the physical objects responsible for initiating and receiving the associated information flow, which includes the data flow under analysis, are included.
- **Functional Objects and Processes:** Descriptions of the functional objects that initiate or receive the data flow and the processes associated with these functional objects are provided, if known.
- **Security Attributes:** Required or recommended security attributes, such as confidentiality, integrity, availability, authentication, and encryption, are provided, if known.
- **STRIDE-based Threats:** The initially identified threats based on the STRIDE model are included from the MS SDL threat report.

In addition to the basic input, the different LLM-based approaches used in the TraCR-TMF utilize different sets of instructions, queries, examples, etc., in the prompts. The following sections explain these approaches in detail.

#### 1) RAG-BASED APPROACH WITH AN LLM

RAG, first introduced by a group of META AI researchers in 2020 [47], became a widely popular technique in natural language processing for its ability to enhance text generation models. RAG provides a way of incorporating relevant external knowledge from a database or document repository in addition to relying on pre-trained knowledge. To this end, we consider an RAG-based approach to complement the generative power of LLMs with retrieval mechanisms to identify pertinent attack techniques from the MITRE ATT&CK matrix. Fig. 3 presents the architecture of our RAG-enabled LLM-based mapping to the MITRE ATT&CK techniques, comprising the following steps:

```
Data Flow: <data_flow_name>
Initiator: <initiator_name>
Acceptor: <acceptor_name>
Requires Authentication?: …
Requires Encryption?: …

Description of the Initiator: <initiator_description>
Function: <function_name_and_desc>
      <process1>: <description>
      <process2>: <description>
      …

Description of the Acceptor: <acceptor_description>
Function: <function_name_and_desc>
      <process1>: <description>
      <process2>: <description>
      …
Definition of <data_flow_name>: <data_flow_definition>

STRIDE-based threats associated with the data flow:
…
```

**FIGURE 2.** **Basic input for LLM.**

**Initial Identification of Relevant Cyberattacks:** The process initiates with a user input, known as the vanilla prompt, which primarily includes a query to identify the cyberattacks that are potentially relevant to a given data flow. The query provided in the vanilla prompt is as follows: "*What are the possible cyberattacks that can be used to attack this data flow? Return them in a Python dictionary format with the key being the cyberattack technique and the value being the technique description*." In addition to this query, the prompt contains the basic input, which we presented in Fig. 2 before. The vanilla prompt is processed by an LLM, i.e., GPT-4o, which we denote as the LLM agent #1 in Fig. 3. Although GPT-4o is not free of charge, it is optimized for inference-based reasoning [48], making it well-suited for identifying implicit cyberattack techniques in our framework. In addition, the model's faster response time, enhanced coherence, and context retention ability compared to its legacy models [49] motivated us to consider it for the TraCR-TMF. As shown in Fig. 3, the LLM agent #1 serves as an intelligent agent that generates a list of general cyberattack techniques based on the query and the basic input.

**Construction of a Vector Database from the MITRE ATT&CK Matrix:** MITRE ATT&CK descriptions are retrieved from files using a data connector, i.e., SimpleDirectoryReader [50]. The files are then processed in batches to optimize memory usage. Next, the text data is converted into embeddings utilizing one of OpenAI's latest embedding models, i.e., text-embedding-3-small model [51]. This embedding model is chosen due to its cost-effectiveness and balance between speed and accuracy [52]. If an index already exists, new documents are inserted into it incrementally. The indices persist, enabling future use without requiring reprocessing. A VectorIndexRetriever [53] is set up to retrieve relevant embeddings based on similarity. Finally, a SimilarityPostprocessor [54] is applied with a cutoff of 0.6 to filter out low-relevance results [54]. A cutoff of 0.6 was selected as it offers a balanced trade-off between precision and recall. A higher threshold excludes potentially useful but moderately similar entries, reducing recall, especially in cases where relevant descriptions use varied phrasing. Conversely, a lower threshold introduces semantically weak matches, reducing precision and increasing noise. The 0.6 cutoff showed to effectively filter out irrelevant content while retaining diverse yet meaningful results.

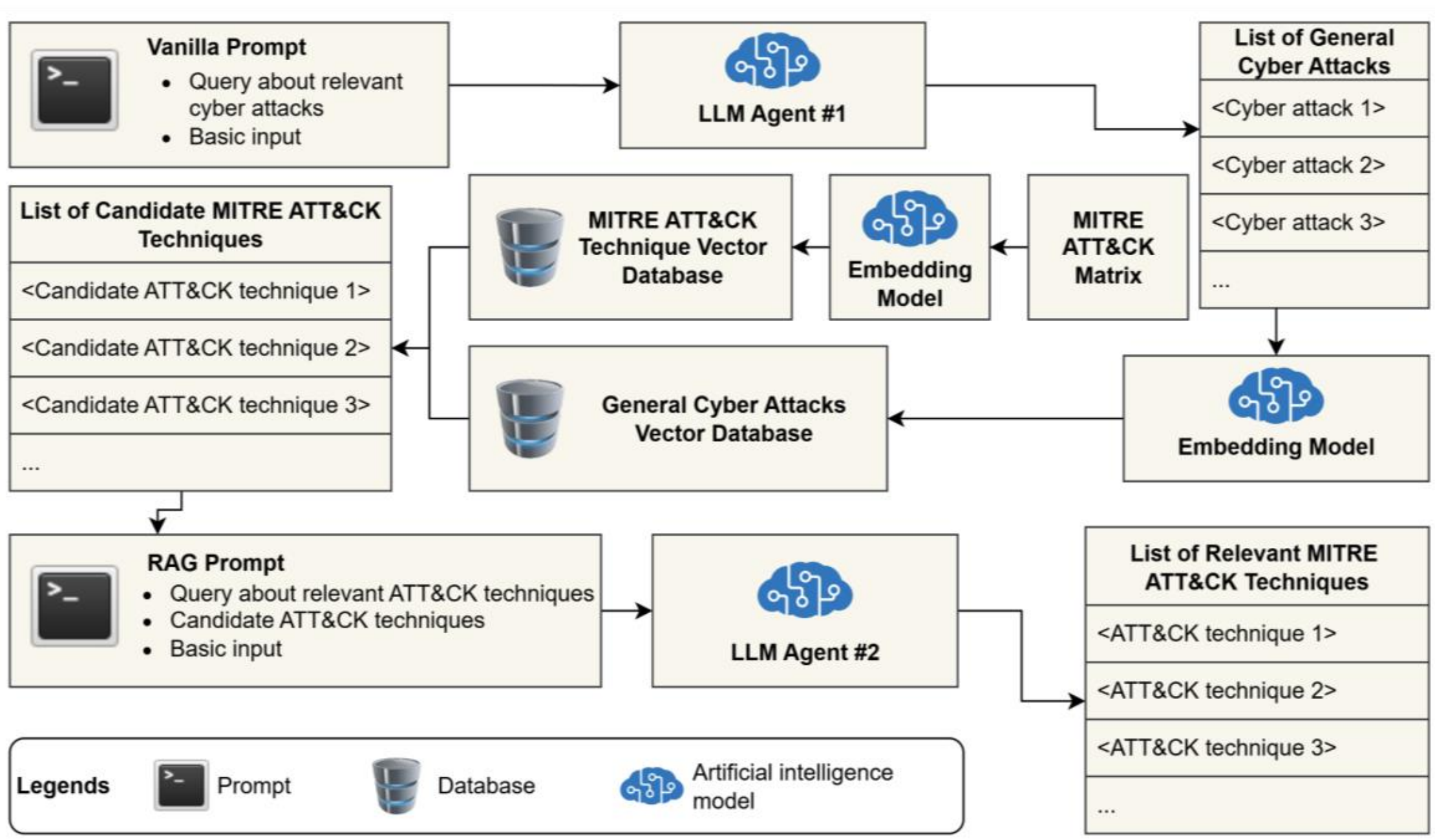

**FIGURE 3.** RAG-based LLM architecture for ATT&CK technique mapping.

**Retrieval of Candidate ATT&CK Techniques from Vector Databases:** The list of general cyberattacks produced by the LLM agent #1 is transformed into vector embeddings using OpenAI's text-embedding-3-small model [51]. These embeddings are then matched against pre-vectorized descriptions of MITRE ATT&CK techniques using cosine similarity to identify the most relevant attack techniques. The top three similar MITRE ATT&CK techniques identified by the LLM agent #1 for each general cyberattack are then selected and compiled into a candidate list, as depicted in Fig. 3.

**Identification of ATT&CK Techniques using a RAG Prompt:** The candidate MITRE ATT&CK techniques are utilized to create a new prompt, which we refer to as the RAG prompt, as shown in Fig. 3. This prompt incorporates the candidate technique IDs, names, and descriptions retrieved from the MITRE ATT&CK matrix. It is followed by the query: "*Which MITRE ATT&CK techniques from the table above can be used to attack the data flow? Provide the technique IDs in Python list format*." Additionally, the RAG prompt includes descriptions of the associated data flow, the initiator and the acceptor of the data flow to enrich the context. The RAG prompt is then processed by another LLM agent, which we refer to as the LLM agent #2 (also based on GPT-4o). LLM agent #2 refines the information and generates a final list of specific MITRE ATT&CK techniques relevant to the data flow. Again, the GPT-4o model is chosen for this task due to its superior logical reasoning compared to OpenAI's legacy models [49].

The RAG-based approach leverages the complementary strengths of generative and retrieval-based approaches, enabling contextually informed predictions of relevant MITRE ATT&CK techniques. Unlike the other two approaches we present in the following sections, this RAG-based approach identifies relevant MITRE ATT&CK techniques without any specific examples or retraining that would require a human cybersecurity expert's intervention. Therefore, this approach is the most automated, requiring no cybersecurity expert intervention for relevant attack technique identification among the three approaches offered by the TraCR-TMF. In addition, this RAG-based approach enables the retrieval of relevant information from the most up-to-date list of attack techniques available in the MITRE ATT&CK matrix at the time of retrieval. Consequently, the information retrieved by this approach is not limited to the training cutoff period, as is the case with the supervised fine-tuning approach.

### 2) IN-CONTEXT LEARNING APPROACH WITH AN LLM

In-context learning (ICL) is a technique that leverages the capabilities of LLMs to perform tasks without requiring retraining. Instead, by providing examples and instructions directly within the prompt, ICL enables an LLM to generalize and produce more accurate results for new inputs. We consider the gpt-4o-mini-2024-07-18 model with ICL to identify the MITRE ATT&CK techniques that are potentially relevant to a given transportation CPS data flow. Three variations of ICL are considered in this context: (i) zero-shot, (ii) one-shot, and (iii) few-shot learning.

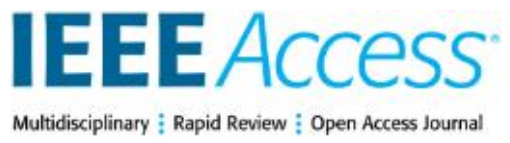

In zero-shot learning, the LLM is provided with only task instructions along with the basic input shown in Fig. 2, without any examples. This approach utilizes the model's ability to generalize solely based on its pre-trained knowledge and the given context. Thus, zero-shot learning is equivalent to using the basic input without any examples.

In one-shot learning, the LLM is given a single example of data flow along with its corresponding MITRE ATT&CK techniques identified by human cybersecurity experts. This example provides the model with a reference for understanding the task and the expected output format.

In a few-shot learning, the LLM is supplied with multiple examples of data flows and their corresponding MITRE ATT&CK techniques identified by human cybersecurity experts. By exposing the model to various instances, this approach improved the LLM's comprehension of the task and enhanced its ability to generalize.

Fig. 4 illustrates the ICL prompt given to the LLM. The prompt is designed to ensure that the LLM receives adequate context to perform the requested task. Since the ICL-based approach includes a few worked-out examples verified by cybersecurity experts, this approach is considered to require low cybersecurity expert intervention.

### 3) SUPERVISED FINE-TUNING APPROACH WITH AN LLM

The supervised learning approach utilizes the power of pre-trained transformer-based models. Due to the complexity and lengthiness of our input texts, we selected ModernBERT over BERT. ModernBERT is a cutting-edge, encoder-only model that was trained on a wide range of English texts. Unlike the original BERT, it can handle longer input sequences [55], which is important in our case as the longest input sequence in our dataset requires nearly seven times BERT's 512-token limit. The base version of ModernBERT, with its 22 layers and 149 million parameters, is chosen for this task since ModernBERT has been shown to strike a balance between strong performance and computational demands [55]. Unlike other state-of-the-art LLMs with superior reasoning capability, such as GPT-4o, ModernBERT is open source and can be downloaded and run on a local machine. All these advantages of using ModernBERT make it an appropriate candidate for our attack technique identification task.

To perform supervised fine-tuning of the ModernBERT model, we follow a transfer learning approach, i.e., we start with a pre-trained model and fine-tune it for the task of our interest. The primary advantage of transfer learning in this case over training a model from scratch is that by leveraging the knowledge already embedded in the pre-trained model from its initial training on large online datasets, the model does not need to learn basic features from the ground up, which yields an improved performance especially when the target dataset is comparatively smaller.

In machine learning, supervised methods can be used when the training dataset contains ground-truth labels. In a multi-label classification problem, each instance can be associated with more than one label. Consider, we denote the dataset as follows: $D_{train} = \{(\mathbf{x_1}, \mathbf{y_1}), (\mathbf{x_2}, \mathbf{y_2}), (\mathbf{x_3}, \mathbf{y_3}), \dots, (\mathbf{x_N}, \mathbf{y_N})\}$, where $\mathbf{x_i} \in \mathbb{R}^N$ is the representation of the $i^{th}$ instance, and $\mathbf{y_i} = [y_{i,1}, y_{i,2}, y_{i,3}, \dots, y_{i,K}]$ is a binary vector with $y_{i,j} = 1$ indicating the $j^{th}$ label is associated with the $i^{th}$ instance based on the ground-truth list, and $y_{i,j} = 0$ indicating otherwise. Then, our goal is to train a multi-label classifier based on ModernBERT that can perform the mapping from $\mathbf{x_i}$ to $\mathbf{y_i}$. In the context of mapping a basic input associated with a transportation CPS data flow to a list of relevant MITRE ATT&CK techniques, $\mathbf{x_i}$ represents the $i^{th}$ basic input and $\mathbf{y_i}$ represents the relevance of the MITRE ATT&CK techniques present in the ground-truth list. The binary cross-entropy loss function considered for retraining the ModernBERT model is given by,

$$L = \frac{1}{NK} \sum_{i=1}^{N} \sum_{j=1}^{K} [y_{i,j} \log(\hat{y}_{i,j}) + (1 - y_{i,j}) \log(1 - \hat{y}_{i,j})] \quad (1)$$

where $\hat{y}_{i,j}$ is the probability assigned by the classifier for the $j^{th}$ MITRE ATT&CK technique to be associated with the $i^{th}$ data flow. To enhance the model's robustness and generalizability, a five-fold cross-validation is performed. We utilize the MultilabelStratifiedKFold method to prepare the train and test splits (80% and 20%, respectively), ensuring balanced label distributions across folds in this multi-label setting. For each fold, a validation set is randomly selected from the corresponding training set. The model is fine-tuned using the Trainer API from the Hugging Face Transformers library (version 4.54.0.dev0) [56]. All layers of the model are updated during this process, employing the default AdamW optimizer. Early stopping is incorporated to prevent overfitting of the training data by monitoring the validation loss with a tolerance of 5 epochs and a minimum improvement threshold of 0.001 (i.e., training stops if the validation loss does not decrease by at least 0.001 for 5 consecutive epochs). The model is configured with a maximum of 50 epochs, a batch size of 8, a learning rate of 0.00002, and a weight decay of 0.01. The batch size of 8 was selected to ensure all input representations could be processed within the memory capacity of an A100 GPU, while the learning rate and weight decay were empirically tuned to optimize model performance. The binary classification threshold for each label (i.e., each MITRE ATT&CK technique present in the training set) can be optimized to yield the maximum F1 scores on the validation sets with the algorithm presented in APPENDIX A.

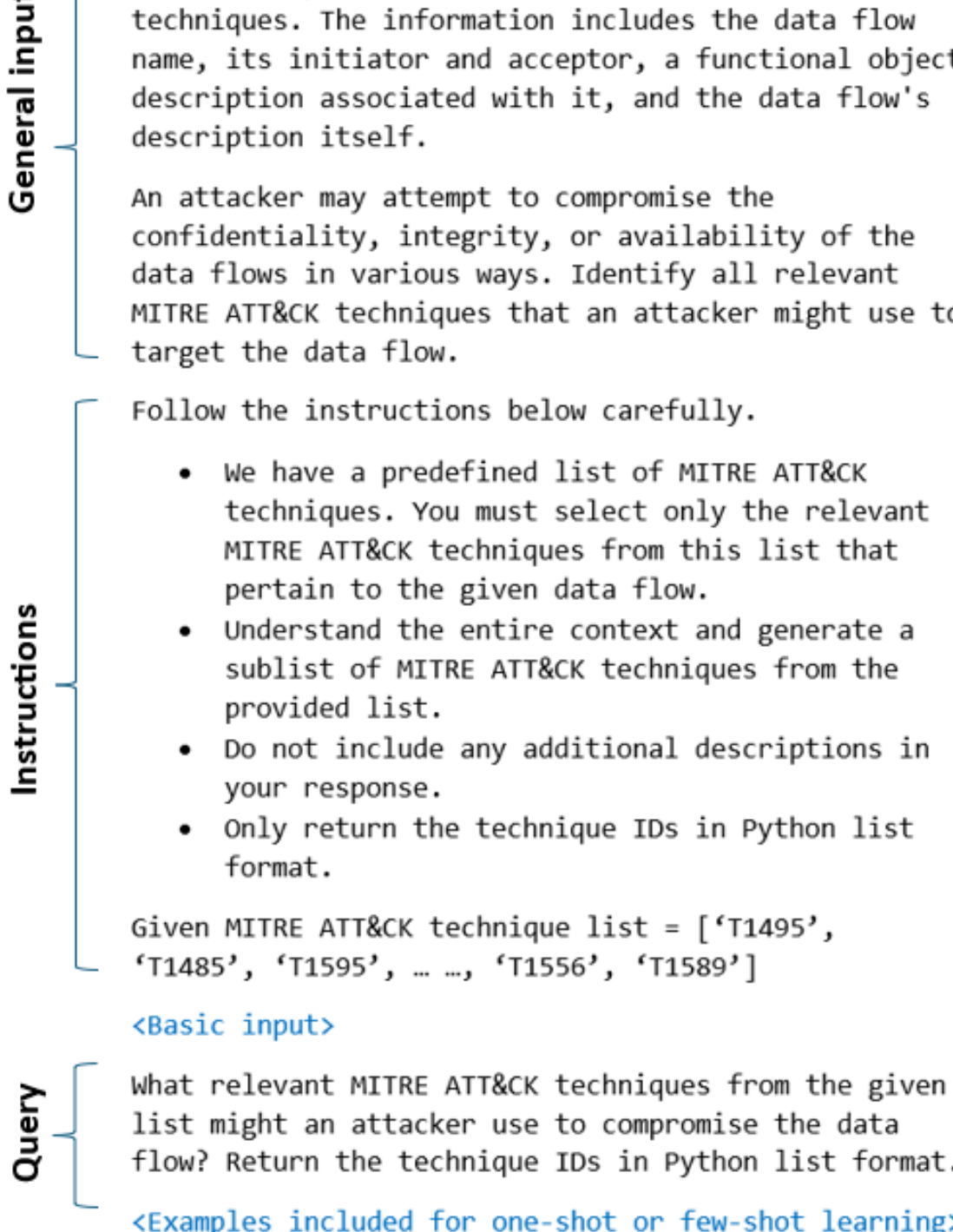

**FIGURE 4. Prompt for LLM while using ICL.**

### C. ASSET-CENTRIC THREAT MODELING USING A CUSTOMIZED LLM (STAGE 3)

Asset-centric threat modeling focuses on identifying threats associated with specified assets that could lead the assets to be compromised. These assets can be physical or functional objects, data stores, or data flows in a transportation CPS application. This type of modeling helps cybersecurity professionals realize potential attack paths exploiting multi-layered vulnerabilities in a transportation CPS application, and associated attack methods or tools that could be utilized by attackers to compromise an organization's valuable assets. Predicting these attack paths helps identify proactive and reactive security measures that need to be incorporated to minimize associated risks and impacts. To this end, the TraCR-TMF offers an asset-centric threat modeling approach to predict potential attack paths leading to compromising critical assets. In this context, an attack path can consist of multiple steps where an attack propagates from one entity to another within a transportation CPS network until it reaches the target asset that the attacker wants to compromise. This is obtained by analyzing the potential MITRE ATT&CK techniques identified for different data flows in a transportation CPS application using a customized LLM. Each predicted attack path is broken down into individual attack steps that complete the attack path. Here, an attack step within a predicted attack path implies the attack propagating from one entity (e.g., an initiator or an acceptor of a data flow) to another entity. In addition, relevant attack techniques that could be used by attackers to conduct each step of the predicted attack paths are explored. The following sections detail how this asset-centric threat modeling is carried out by an LLM in the TraCR-TMF.

#### 1) BUILDING A CUSTOMIZED GPT WITH CHATGPT

Powered by GPT-4o, OpenAI now offers users the opportunity to build their customized versions of ChatGPT with a single line of instruction [57]. This enables users to build their customized GPT with expertise in specialized tasks. In addition, enabling OpenAI's memory feature helps the customized GPT retain insights over sessions to refine interactions over time. To build our customized GPT for asset-centric threat modeling, we provide the following one-line instruction to ChatGPT: "*Build a customized GPT for me that will serve as an expert-level cybersecurity analyst*." Given this prompt, ChatGPT builds a customized GPT, specialized for cybersecurity analysis.

#### 2) PROMPT CONSTRUCTION FOR ASSET-CENTRIC THREAT MODELING

Once the customized GPT is ready for user interactions, the GPT is provided with a prompt to perform the asset-centric threat modeling. This prompt includes the names of the initiator and the acceptor of each data flow, along with their associated MITRE ATT&CK techniques, in a table format, as shown in Fig. 5. This information is followed by a set of instructions and a query, asking the customized LLM to identify potential attack paths leading to a specified asset to be compromised along with the associated MITRE ATT&CK techniques that can be utilized in each step of the attack path.

As shown in Fig. 5, the customized GPT is also instructed to present its output in a table format with one column showing the attack paths it predicted that could compromise the specified asset, and another column explaining the attack execution steps with associated MITRE ATT&CK techniques. Thus, the output from the customized GPT provides users with asset-centric threats that would help cybersecurity experts think critically about how such attacks could be detected, mitigated, or prevented in the first place.

### D. THREAT MITIGATION PRIORITIZATION (STAGE 4)

To prioritize the mitigation of identified threats, which is a common practice among cybersecurity professionals given the vast number of existing vulnerabilities, the TraCR-TMF leverages CVSS scores. As mentioned in Section IV-E, CVSS is a widely accepted open framework for assessing the severity of vulnerabilities [16]. The NVD reports the CVSS base scores for CVEs [46]. The Prioritize Known Exploited Vulnerabilities program [58], launched by The MITRE Corporation's Center for Threat-Informed Defense (CTID), mapped the previously exploited CVEs listed in the Known Exploited Vulnerabilities (KEV) Catalog [15] to relevant MITRE ATT&CK techniques. For each CVE in the KEV Catalog, this mapping identifies associated ATT&CK techniques at three levels: (i) Exploitation technique, which includes ATT&CK techniques used to exploit the vulnerability; (ii) Primary impact, which includes techniques

**General input**

The table below presents potential MITRE ATT&CK techniques for different data flows in a transportation cyber-physical systems application. The first column of each row presents the name of the initiator, and the second column presents the name of the acceptor of a data flow. The third column presents a list of MITRE ATT&CK techniques an attacker could utilize to compromise the data flow.

| Initiator | Acceptor | MITRE ATT&CK Technique ID |
|---|---|---|
| <initiator1> | <acceptor1> | <ID_list1> |
| <initiator2> | <acceptor2> | <ID_list2> |
| <initiator3> | <acceptor3> | <ID_list3> |
| … | … | … |

**Instructions**

Using the information provided in the table above, identify different attack paths from <first_asset_name> to <target_asset_name> that an attacker could take to compromise the <target_asset_name>. Please follow the instructions below carefully:

1. An attack path can include multiple steps. Each step is basically the attack propagating from one entity to another entity. Both entities for a step should always directly come from the initiator and the acceptor present in the same row of the given table.
2. While determining the potential attack paths and the steps included in the attack paths, you cannot use any assets outside the initiators and the acceptors mentioned in the table above.
3. For each step of an attack path, you must determine the potential MITRE ATT&CK techniques from the ones mentioned in the corresponding row of the table. The MITRE ATT&CK techniques you include in each step must be techniques that an attacker could use to complete that step.

The output should be formatted in the following way:

1. The output must be in table format in which you'll have two columns. The first column should present all the steps in an attack path using arrows.
2. The second column should present a step-by-step explanation of the attack path along with the associated MITRE ATT&CK techniques

**Query**

Could you provide me with all the different attack paths according to the above instructions?

**FIGURE 5. Prompt for asset-centric threat modeling.**

reflecting the immediate benefits of exploitation; and (iii) Secondary impact, which includes techniques representing the benefits derived from the primary impact. Details of this mapping method can be found in [59]. The TraCR-TMF uses this mapping, available through CTID's Mappings Explorer [60], to identify the CVE IDs associated with each ATT&CK technique identified using LLMs in stage 2 (explained in Section V-B).

Once the CVEs associated with each relevant ATT&CK technique are identified, the NVD is used to retrieve their CVSS base scores, which reflect both the exploitability and the exploitation impact of the vulnerabilities. A higher CVSS score indicates a more severe vulnerability. Therefore, the higher the CVSS score is, the higher the priority for its mitigation should be. After collecting the CVSS base scores for all associated vulnerabilities, we rank them in the descending order of their CVSS scores. This ranking can then assist cybersecurity professionals in making informed threat mitigation decisions. Note that CVEs are often tied to specific vendors and their products, such as operating systems, libraries, and protocols. As a result, cybersecurity professionals working with a particular transportation CPS application would likely need to narrow down the most relevant CVE IDs from those identified by the TraCR-TMF. This step cannot be fully automated, as it requires evaluating the applicability of each vulnerability based on the specific software and hardware deployed (or planned to be deployed) within each transportation CPS environment, which may vary across jurisdictions even for the same transportation CPS application.

Another potential advantage of mapping the identified threats to known exploited vulnerabilities is that the NVD provides references to related advisories from vendors and third parties. These advisories often include software updates and patches that are essential for threat mitigation, as well as references to available exploit codes, which can support red teaming and penetration testing activities.

## VI. EVALUATION OF TRACR-TMF

To evaluate the TraCR-TMF, we consider two evaluation cases as follows: (i) the evaluation of ARC-IT transportation CPS applications, and (ii) the evaluation of a major real-world cyberattack incident. The first evaluation focuses on identifying the potential MITRE ATT&CK techniques that could exploit the vulnerabilities of different transportation CPS applications using the applications' planning-level systems architecture. In contrast, the second evaluation, i.e., evaluation of a major real-world cyberattack incident, focuses on asset-centric threat modeling to identify potential attack paths leading to compromising a critical transportation CPS asset, in addition to the identification of potential MITRE ATT&CK techniques that could exploit the vulnerabilities. As the real-world cyberattack incident has already taken place and the compromised assets are known for the incident, the predicted attack paths leading to the compromised are verifiable. However, this verification is not possible with the first evaluation case that focuses only on the planning-level transportation CPS applications present in ARC-IT. In addition, the LLMs, fine-tuned or customized for the ARC-IT transportation CPS application evaluation, are directly applied without any further modifications to evaluate the real-world incident considered in this study. This provides an assessment of the TraCR-TMF's transferability beyond what the framework's constituent LLMs are preapred for.

### A. EVALUATION METRICS

To evaluate the performance of the TraCR-TMF when we utilize LLMs to identify attack techniques that are relevant to a given transportation CPS data flow, we consider three standard metrics of classification tasks: (i) precision, (ii) recall, and (iii) F1-score. For a binary classification task, these metrics are calculated as follows:

$$Precision = \frac{TP}{TP + FP} \tag{2}$$

$$Recall = \frac{TP}{TP + FN} \tag{3}$$

$$F1 = \frac{2 \cdot Precision \cdot Recall}{Precision + Recall} \quad (4)$$

where, $TP$ represents the total number of true positives across all labels, $FP$ represents the total number of false positives across all labels, and $FN$ represents the total number of false negatives across all labels. Micro precision and recall measure the proportion of correctly predicted positive instances among all predicted positives (aggregated globally), and among all actual positives (aggregated globally), respectively. The micro F1 score is the harmonic mean of micro precision and recall.

To further clarify how these metrics are calculated for a multi-label classification task, let us consider that $P_i$ denotes the predicted set, $G_i$ denotes the ground-truth set, and $O_i$ denotes the overlap between $P_i$ and $G_i$. Then, precision and recall are given as follows:

$$Precision = \frac{1}{N}\sum_{i=1}^{N}\frac{|O_i|}{|P_i|} \quad (5)$$

$$Recall = \frac{1}{N}\sum_{i=1}^{N}\frac{|O_i|}{|G_i|} \quad (6)$$

where, $N$ is the total number of data flows considered.

### *B. EVALUATION OF ARC-IT TRANSPORTATION CPS APPLICATIONS*

The objective of this evaluation is to assess the TraCR-TMF's efficacy in identifying the relevant attack techniques and corresponding countermeasures by applying the framework to a wide range of transportation CPS applications available on ARC-IT. To this end, we consider 26 different randomly selected transportation CPS applications from ARC-IT [45]. In total, we extracted 433 different data flows that are related to these transportation CPS applications along with their associated initiator, acceptor, functional objects and processes, and recommended security attribute information, such as confidentiality, integrity, availability, encryption, and authentication requirements. This information helps us construct the basic input, shown in Fig. 2, except for the part dedicated to STRIDE-based associated threats that come from the MS SDL threat modeling tool.

ARC-IT serves as a database for the U.S. national reference architectures for transportation CPS applications that are widely adopted by transportation professionals for implementation. This database provides detailed physical views of diverse transportation CPS application architectures, outlining different involved entities and their interactions. These reference architectures provide adequate information to develop their corresponding DFDs in the MS SDL threat modeling tool. Once a transportation CPS application architecture is transferred to the MS SDL threat modeling tool, the tool automatically identifies potential threats for each data flow within the application according to the STRIDE model. Fig. 6 presents a snapshot of the threat report generated by the MS SDL tool for an example transportation CPS application that is referred to as the "CVO03: Electronic Clearance" service package in ARC-IT [61]. The threat information obtained from the MS SDL threat modeling report completes the basic input (shown in Fig. 2) requirements for each data flow.

The basic input is then coupled with instructions and queries to prepare the prompts to be processed by the LLMs to identify relevant attack techniques from the MITRE ATT&CK matrix. As mentioned in Section V-B, TraCR-TMF utilizes three different LLM-based approaches to map the STRIDE-based threats to potentially relevant attack techniques from the MITRE ATT&CK matrix, as follows: (i) a RAG-based approach, (ii) an ICL-based approach, and (iii) a supervised fine-tuning approach. Unlike the RAG-based approach, which does not require a predetermined list of relevant attack techniques, the ICL-based approach requires a few predetermined relevant attack techniques to be provided as examples. On the other hand, the supervised fine-tuning approach requires a ground-truth list of relevant attack techniques to train on. Preparing this predetermined or ground-truth list of relevant attack techniques for each data flow is a task that can only be done by cybersecurity experts. In this study, we established this list by consulting with two cybersecurity experts with decades of academic and industrial experience in the cybersecurity domain. Nevertheless, this list remains incomplete since the MITRE ATT&CK matrix includes over 200 different attack techniques, all of which could not be reviewed by the cybersecurity experts for each of the 433 data flows considered in this evaluation. However, unlike traditional supervised machine learning or deep learning models, LLMs are not only able to perform multi-label classification tasks based on their learning from provided examples or supervised training, but LLMs can go beyond the given examples or its training domain due to its original training on a vast amount of data and ability to understand the contexts [62].

Among the three LLM-based ATT&CK technique identification approaches in TraCR-TMF, only the supervised fine-tuning approach requires training of the model. Fig. 7 shows the training and validation losses from our five-fold cross-validation. For each fold, the solid line represents the training loss, and the dashed line of the same color represents the corresponding validation loss. For each fold, the training was terminated based on the stopping criteria discussed in Section V-B-3.

The LLM-predicted relevant attack techniques are compared with the corresponding ground-truths to determine the precision, recall, and F1 score for each LLM-based approach. Table IV presents these results for the three LLM-based attack technique identification approaches. Note that the results may vary slightly across runs due to minor differences in floating-point operations on Graphics Processing Units

Diagram 1 Diagram Summary:

| | |
|---|---|
| Not Started | 341 |
| Not Applicable | 0 |
| Needs Investigation | 0 |
| Mitigation Implemented | 0 |
| Total | 341 |
| Total Migrated | 0 |

Interaction: (1A) electronic screening request

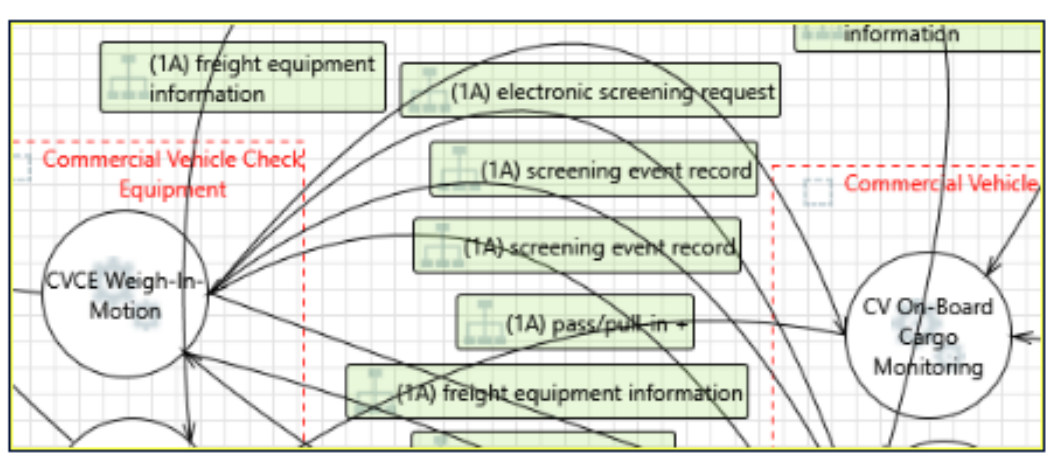

1. Data Flow Sniffing [State: Not Started] [Priority: High]

Category: Information Disclosure
Description: Data flowing across (1A) electronic screening request may be sniffed by an attacker. Depending on what type of data an attacker can read, it may be used to attack other parts of the system or simply be a disclosure of information leading to compliance violations. Consider encrypting the data flow.
Justification: <no mitigation provided>

2. Potential Data Repudiation by CV On-Board Cargo Monitoring [State: Not Started] [Priority: High]

Category: Repudiation
Description: CV On-Board Cargo Monitoring claims that it did not receive data from a source outside the trust boundary. Consider using logging or auditing to record the source, time, and summary of the received data.
Justification: <no mitigation provided>

3. Potential Lack of Input Validation for CV On-Board Cargo Monitoring [State: Not Started] [Priority: High]

Category: Tampering
Description: Data flowing across (1A) electronic screening request may be tampered with by an attacker. This may lead to a denial of service attack against CV On-Board Cargo Monitoring or an elevation of privilege attack against CV On-Board Cargo Monitoring or an information disclosure by CV On-Board Cargo Monitoring. Failure to verify that input is as expected is a root cause of a very large number of exploitable issues. Consider all paths and the way they handle data. Verify that all input is verified for correctness using an approved list input validation approach.
Justification: <no mitigation provided>

**FIGURE 6. Snapshot of the STRIDE-based threat report of "CVO03: Electronic Clearance" application.**

(GPUs), which can influence the probabilistic identification of attack techniques by the LLMs (see [63]). To enhance reproducibility using the shared codes in our GitHub repository [64], we incorporated specific PyTorch commands to reduce sources of nondeterministic behavior. However, in practical deployments, it may be beneficial to consider all relevant attack techniques identified across multiple runs, which can be obtained by disabling these commands.

As observed from Table IV, the supervised fine-tuning approach outperformed the RAG-based and the ICL-based attack technique identification approaches. While the RAG-based approach performed the worst, it should be noted that this approach solely relied on the basic input and utilized its multistage attack identification architecture, as shown in Fig. 3, to identify relevant attack techniques from the MITRE ATT&CK matrix. In contrast, the ICL-based approach leveraged in-prompt examples to learn relevant attack techniques and the supervised fine-tuning approach leveraged retraining the pre-trained ModernBERT LLM to enhance the model's attack technique identification performance. Among zero-, one-, and few-shot ICL approaches, the few-shot learning performed the best. While we tested few-shot learning by providing two to nine examples within a single prompt, few-shot learning with eight examples outperformed the other ICL approaches, which is reported in Table IV as few-shot learning. This better performance of eight example-based few-shot learning can be attributed to striking the right balance between better learning from additional examples and potential hallucinations due to a context overload issue induced by lengthy prompts with excess examples [65]. Although the supervised fine-tuning approach outperformed the best-performing ICL-based approach in this evaluation, fine-tuning requires establishing ground truths, which could be challenging to obtain with limited intervention from cybersecurity experts. In contrast, providing a handful of examples to leverage ICL might seem easier for potential adoption in some cases.

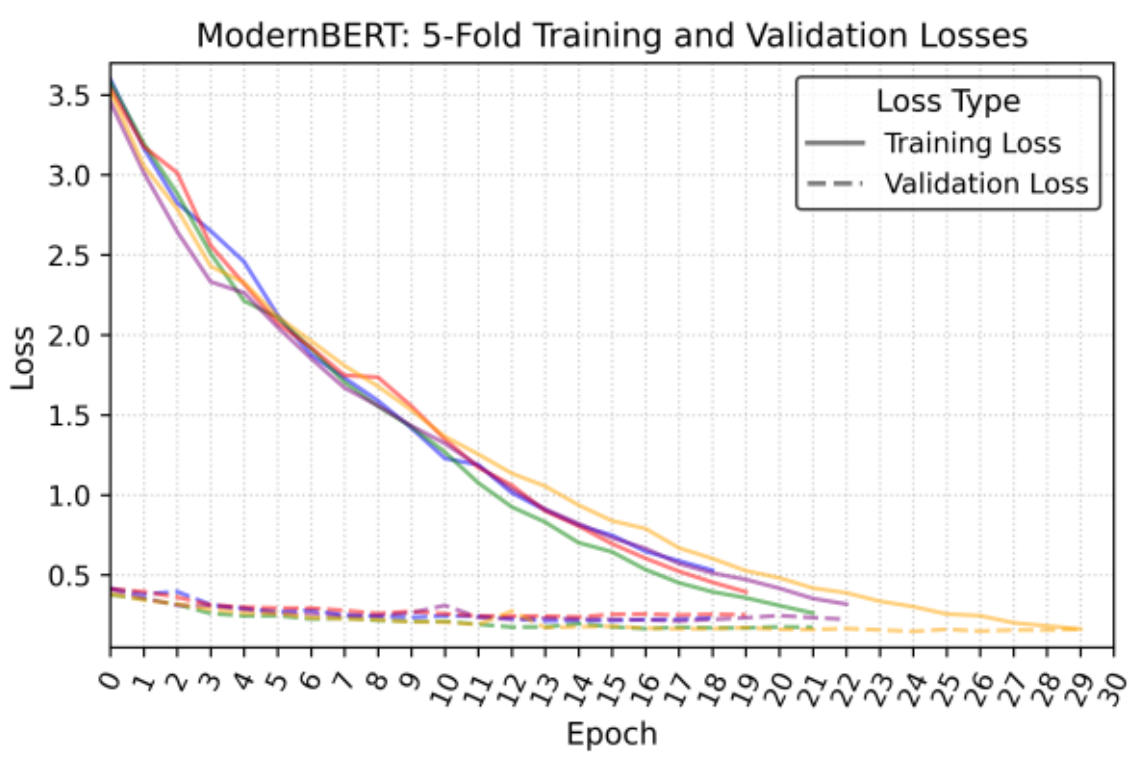

**FIGURE 7. Training and validation losses during LLM fine-tuning.**

TABLE IV
PERFORMANCE OF TRACR-TMF IN IDENTIFYING RELEVANT ATT&CK TECHNIQUES

| LLM-based Approach | | Precision | Recall | F1 Score |
|---|---|---|---|---|
| RAG | | 0.33 | 0.20 | 0.25 |
| ICL | Zero-shot | 0.18 | 0.18 | 0.18 |
| | One-shot | 0.38 | 0.31 | 0.34 |
| | Few-shot | 0.47 | 0.48 | 0.48 |
| Supervised fine-tuning | | 0.70 | 0.75 | 0.72 |

Another insight into the performance of the LLM-based relevant attack identification approaches is that the LLMs identified several attack techniques outside their training domain, several of which were found to be relevant. As mentioned earlier, the ground-truth list is not complete. The cybersecurity experts, whom we consulted to develop the ground-truth list in this study, reviewed several, but not all, of the existing MITRE ATT&CK techniques. Even if we had established a ground-truth list by considering every attack technique from the MITRE ATT&CK matrix, new attack techniques would be introduced and added to the matrix in the future. Thus, LLMs' ability to identify relevant attack techniques beyond their training domain is useful.

To better assess LLM's performance in identifying relevant attack techniques from the MITRE ATT&CK matrix, we randomly selected 50 data flows out of the 433 data flows used in this evaluation and again consulted with the cybersecurity experts to evaluate each LLM-predicted attack technique that was not present in our ground-truth list of relevant attack techniques. Based on this second stage validation from the cybersecurity experts who participated in this study, we reevaluated the performance of the supervised fine-tuning approach on the randomly selected 50 data flows, which is presented in Table V. In the table, it is observed that once we updated the ground-truth list with the help of the cybersecurity experts, considering all the attack techniques identified by the supervised LLM, the precision improved from 0.64 to 0.73 (i.e., a 14% improvement in precision) for the randomly selected 50 data flows. This indicates that approximately 73% of the attack techniques identified by the supervised LLM were correct. This helps us consider the supervised LLM-based approach for our second evaluation case related to a major real-world cyber incident.

TABLE V
PERFORMANCE OF SUPERVISED LLM ON 50 RANDOMLY SELECTED TRANSPORTATION CPS DATA FLOWS

| Supervised LLM-based Approach | Precision | Recall | F1 Score |
|---|---|---|---|
| Based on initial ground truths set by the experts | 0.64 | 0.67 | 0.65 |
| Based on the LLM-identified attack techniques with subsequent expert validation | 0.73 | 0.66 | 0.69 |

Fig. 8 presents two bar charts showing the top 50% of the MITRE ATT&CK techniques identified as false positives (in Fig. 8a) and false negatives (in Fig. 8b) across the 50 randomly selected data flows. Here, false positives indicate cases where the supervised LLM correctly predicted relevant ATT&CK techniques outside the ground-truth list, which was validated by the cybersecurity experts who participated in this study, while false negatives reflect missed predictions of techniques that were present in the ground-truth list. Fig. 8a (presenting the false positives) shows that the supervised LLM can identify a broader range of ATT&CK techniques than it was exposed to during fine-tuning. For example, Fig. 8a shows that T1136 (Create Account) [66] was identified by our supervised LLM as relevant for 11 out of the 50 data flows, all of which were confirmed as correct by the cybersecurity experts who participated in this study. This capability can be attributed to the model's extensive prior knowledge and contextual reasoning abilities. In contrast, Fig. 8b (presenting the false negatives) shows that the model consistently missed some relevant ATT&CK techniques. For example, T1098 (Account Manipulation) [67] was missed by our supervised LLM for 13 out of the 50 data flows. In practice, if cybersecurity professionals consider any of these missed techniques critical to their use case, they can adjust the corresponding thresholds of those techniques, thereby adjusting the true positives or the false positives, using the algorithm in APPENDIX A.

Stage 4 of our TraCR-TMF (explained in Section V-D) helps prioritize threat mitigations based on the severity of associated vulnerabilities with known exploitation histories. To illustrate how this stage supports cybersecurity professionals, consider the Traffic Detector Coordination data flow in a TM16: Reversible Lane Management application [68]. First, we identified the CVE IDs of known exploited vulnerabilities associated with each LLM-identified MITRE ATT&CK technique for this data flow using CTID's Mappings Explorer [60]. Second, we used the CVSS Version 3.1 base scores from the NVD [46] to rank the vulnerabilities. CVSS Version 3.1, i.e., the immediate predecessor to the latest Version 4.0, is considered here because not all vulnerabilities listed in the KEV Catalog had CVSS Version 4.0 scores available at the time of this study. Table VI presents the vulnerabilities associated with ATT&CK techniques relevant to the Traffic Detector Coordination data flow, ranked in descending order of their CVSS scores. Vector strings, showing how these scores were calculated, are available in the NVD. As shown in Table VI, T1190 (Exploit Public-Facing Application) [69] is associated with four different known exploited vulnerabilities, each with a CVSS base score of 10. Given the severity of these critical vulnerabilities, cybersecurity professionals working with a TM16 application [68] may prioritize mitigating these threats before addressing the other vulnerabilities of lower severity in Table VI.

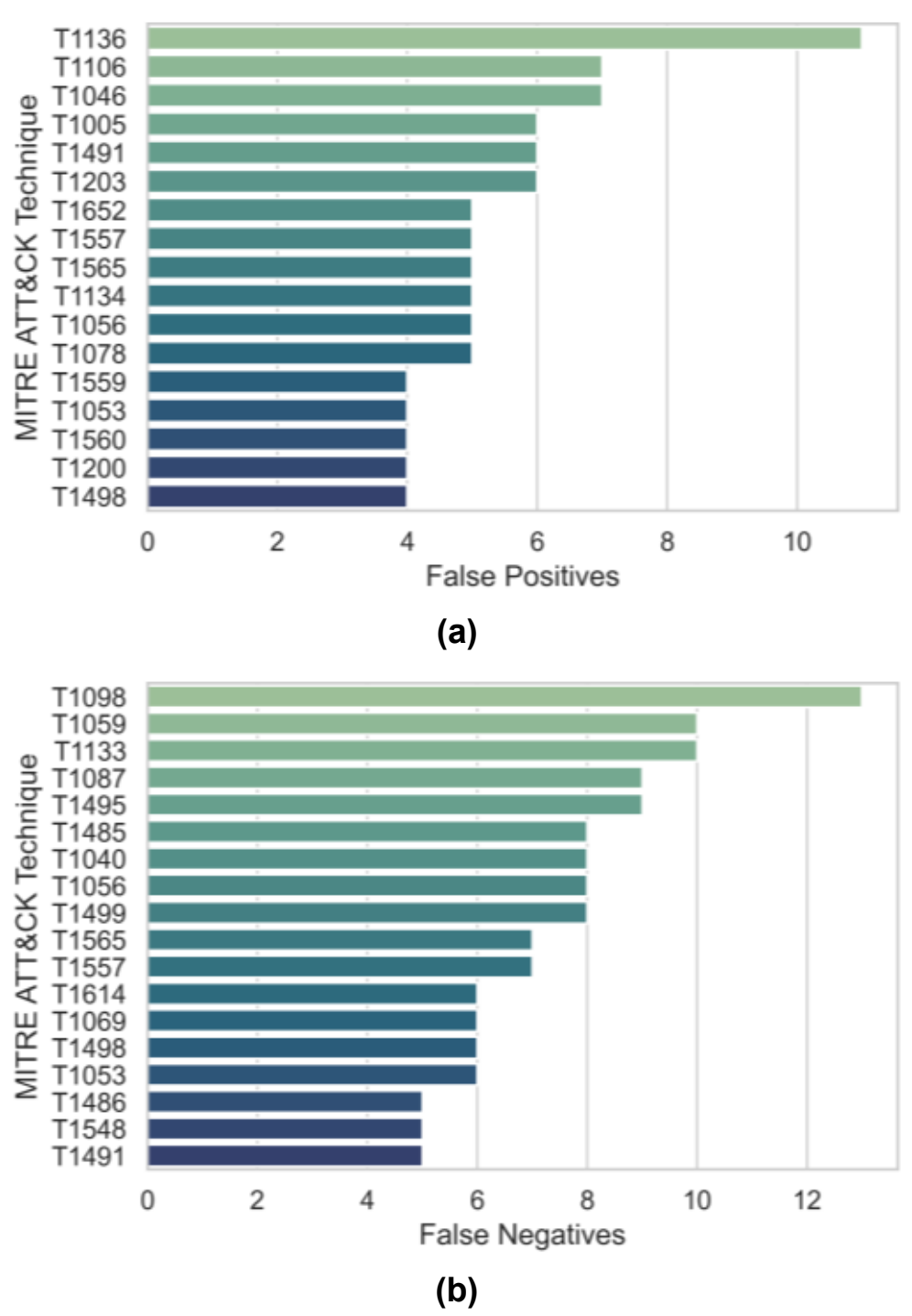

**FIGURE 8.** **(a) False positives and (b) false negatives for 50 randomly selected transportation CPS data flows.**

Note that not all identified vulnerabilities may apply to a deployed or planned TM16 application. For instance, consider the first vulnerability listed in Table VI associated with T1190, i.e., CVE-2021-44228, which was found in the Java Naming and Directory Interface (JNDI) features of Apache Log4j2 versions 2.0-beta9 through 2.15.0 [70]. Whether this applies depends on the use of these logging framework versions in a TM16 application. If deemed applicable, the corresponding advisories, solutions, or tools can be found on the NVD webpage (see [70]). Thus, in addition to the detection and mitigation strategies for techniques listed in the MITRE ATT&CK matrix, stage 4 of our framework helps identify vulnerability-specific mitigations.

TABLE VI
ATT&CK TECHNIQUES RANKED BY CVSS SCORES OF ASSOCIATED CVE IDS FOR TRAFFIC DETECTOR COORDINATION DATA FLOW

| Rank | ATT&CK Techniques | Associated CVE IDs | CVSS Version 3.1 Base Score |
|---|---|---|---|
| 1 | T1190 (Exploit Public-Facing Application) | CVE-2021-44228<br>CVE-2021-22205<br>CVE-2023-20198<br>CVE-2022-22947 | 10 (Critical) |
| 2 | T1040 (Network Sniffing) | CVE-2022-1040 | 9.8 (Critical) |
| | T1059 (Command and Scripting Interpreter) | CVE-2022-35405<br>CVE-2024-4879<br>CVE-2024-5217<br>CVE-2022-35914 | 9.8 (Critical) |
| | T1105 (Ingress Tool Transfer) | CVE-2015-5119<br>CVE-2016-1019<br>CVE-2016-4117<br>CVE-2010-2861<br>CVE-2023-38203<br>CVE-2023-29300<br>CVE-2018-15982<br>CVE-2021-44515<br>CVE-2023-3519 | 9.8 (Critical) |
| 3 | T1565 (Data Manipulation) | CVE-2021-31207 | 6.6 (Medium) |
| 4 | T1557 (Adversary-in-the-Middle) | CVE-2019-5591 | 6.5 (Medium) |

### C. EVALUATION OF A MAJOR REAL-WORLD CYBERATTACK INCIDENT

The objectives of this real-world cyberattack incident evaluation are as follows: (i) to assess the transferability of the TraCR-TMF by applying its underlying supervised LLM to transportation CPS outside the model's training domain, (ii) to perform an asset-centric threat modeling of transportation CPS using the TraCR-TMF to identify potential attack paths and associated attack techniques leading to a target asset, (iii) to assess the attack paths and the associated attack techniques identified by the framework by verifying them against the known facts of the real-world cyberattack incident. To this end, we selected the Colonial Pipeline double extortion ransomware attack, a major real-world cyberattack incident that took place in 2021.

The Colonial Pipeline incident serves as an evaluation case for which we can assess the attack paths and the associated techniques identified by the TraCR-TMF. To perform an asset-centric threat modeling using our framework, a target asset must be specified for which we aim to identify (i) the potential attack paths leading to that asset, and (ii) the associated attack techniques for each exploitation involved in those attacks. Ideally, if the framework could accurately predict the potential attack paths and associated attack techniques, an attacker could utilize any of those means to compromise the target asset. Choosing a widely discussed real-world cyberattack incident, like the Colonial Pipeline, is useful for this evaluation since different public and private security organizations have published reports highlighting their insights related to the incident, helping us better understand the actual cyberattack incident and compare the predictions of our framework with the actual incident. In addition, to identify the relevant attack techniques for each data flow, the LLM fine-tuned for our first evaluation case is utilized without any additional training, which enables us to assess the transferability of the TraCR-TMF. In the following sections, we provide a brief description of how the attack took place to the best of our knowledge, state our assumptions for this evaluation, present the threat modeling evaluation conducted based on the TraCR-TMF, and discuss the evaluation results.

#### 1) COLONIAL PIPELINE RANSOMWARE: KNOWN FACTS

Colonial Pipeline is among the major movers of gasoline and other refined fuels on the East Coast of the U.S. In May 2021, an attacker group utilized the Darkside malware, a ransomware-as-a-service (RaaS) provided by a hacker group, to attack the Colonial Pipeline IT network [71], [72]. About a week after the attack began, the attacker group demanded a $4.4 million ransom in cryptocurrency from the Colonial Pipeline authority. Within this period, the attacker exfiltrated over 100 gigabytes of data outside the Colonial Pipeline's IT network before encrypting the company's valuable data [73].

According to multiple sources [71], [73], [74], the attackers accessed the IT network of the Colonial Pipeline through a compromised virtual private network (VPN) account near the end of April 2021. The account credentials might have been leaked as part of a separate data breach, or the attackers might have blackmailed a legitimate account holder, or the attackers might even guess the credentials, which could not be confirmed based on the publicly available resources. In addition, the VPN account did not have multi-factor authentication (MFA) enabled, which helped the attackers gain access to the VPN by only using the compromised VPN account credentials [73], [74]. Once gained access, the attackers moved laterally within the IT network, gradually exfiltrated over 100 gigabytes of the company's valuable data, and installed ransomware to encrypt the files. Finally, on May 7, the attackers left a note demanding ransom on one of the company's computers [73].

### 2) ASSUMPTIONS FOR COLONIAL PIPELINE EVALUATION

To apply the TraCR-TMF to carry out a threat modeling of the Colonial Pipeline's ICS, we need its systems architecture first. However, the actual architecture of the pipeline's ICS is confidential information that is not publicly available. To this end, we assume that the Colonial Pipeline follows a network architecture similar to the widely known Purdue Model for ICS networks [75] for this evaluation. The Purdue model provides a hierarchical design framework that segregates the software and hardware parts of the network into six different levels, as depicted in Fig. 9. In this architecture, levels 0 through 3 comprise the operational technology (OT) network, and levels 4 through 5 represent the IT network. During the actual ransomware attack, the attackers compromised levels 4 and 5 (i.e., the company's IT network), and there was no evidence indicating the attackers got into the OT network [71]. Once the ransomware incident was identified, the Colonial Pipeline authority disconnected its OT network from the IT network to contain the attack from spreading into the OT network and the systems connected to it [71], [72].

### 3) COLONIAL PIPELINE THREAT MODELING

First, we develop the DFD of the Colonial Pipeline ICS in the MS SDL threat modeling tool based on the Purdue model architecture presented in Fig. 9. Second, the STRIDE-based threat report generated by the MS SDL threat modeling tool is used construct the basic input shown in Fig. 2. Third, the basic input is processed by the supervised LLM trained for the first evaluation case to identify relevant attack techniques from the MITRE ATT&CK matrix. Fourth, the asset-centric threat modeling prompt (shown in Fig. 5) is constructed based on the identified attack techniques. In this prompt, we assign the Business Servers (shown in Fig. 9) as the target asset to be compromised and a human user as the starting point of the attack paths to be identified. Finally, the prompt is processed by the customized GPT model to identify potential attack paths and attack techniques associated with each step of these attack paths that could be utilized by attackers to compromise the Business Servers.

### 4) COLONIAL PIPELINE EVALUATION RESULTS AND DISCUSSIONS

Table VII presents three different potential attack paths and associated attack techniques identified by the customized GPT model in our TraCR-TMF. The model predicted that any of these three attack paths could be followed by an attacker to perform an attack compromising the Business Servers in the Colonial Pipeline IT network. To assess whether these attack paths and their associated attack techniques from the MITRE ATT&CK matrix are feasible, we again consulted the cybersecurity experts who participated in this study. Their expert opinions validating the attack paths and their associated

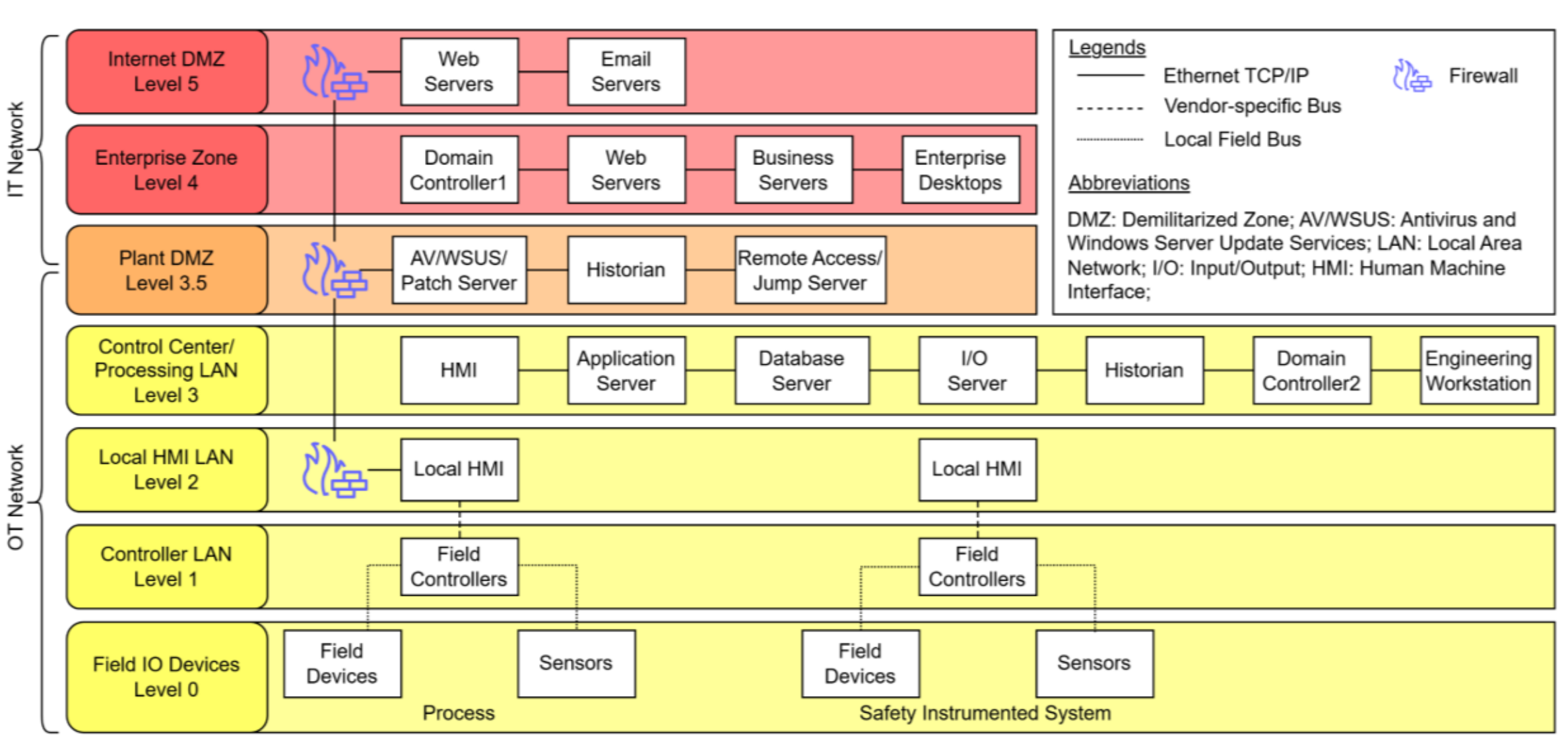

**FIGURE 9. Purdue model for ICS (adopted from [75]).**

attack techniques from the MITRE ATT&CK matrix are summarized below:

**Potential Attack Path #1:** T1552 (Unsecured Credentials) [76] is not a directly relevant technique for this attack path as this technique is about obtaining unsecured stored credentials, which the Colonial Pipeline attacker already possessed. On the other hand, T1059 (Command and Scripting Interpreter) [77] is relevant because the initial access payload must have included some remote execution-related commands and scripts that helped the adversary get within the systems. T1105 (Ingress Tool Transfer) [78] goes beyond maintaining presence and is used to transfer/spread tools between victim devices within a compromised environment; therefore, it is not relevant in this context. On the other hand, T1040 (Network Sniffing) [79] is relevant because it is a basic approach to capture network traffic to realize the network configurations. Similarly, T1557 (Adversary-in-the-Middle) [80] is among the open-source techniques commonly used by the ALPHV Blackcat group [81], believed to be associated with DarkSide, to obtain credentials and session cookies that could help attackers move laterally within a network; therefore, it is a relevant technique for this attack path. T1565 (Data Manipulation) [82] is also relevant because the attackers could delete/modify some critical data after exfiltrating it outside the company's network. However, T1495 (Firmware Corruption) [83] is not relevant as there is no evidence that indicates the attackers performed firmware modification during the Colonial Pipeline incident.

**Potential Attack Path #2:** T1059 (Command and Scripting Interpreter), T1557 (Adversary-in-the-Middle), and T1040 (Network Sniffing) are relevant for this attack path as well, while T1105 (Ingress Tool Transfer) and T1495 (Firmware Corruption) are not relevant, as explained for the first predicted attack path. T1548 (Abuse Elevation Control Mechanism) [84] is relevant, because once the attackers gain access to the Domain Controller1, they could elevate their privileges to help move to the Business Servers. Although T1552 (Unsecured Credentials) is not relevant for step 1 of this attack path (shown in Table VII), it is relevant for step 4 to obtain the credentials for Business Servers. On the other hand, T1020 (Automated Exfiltration) is directly relevant because the attackers exfiltrated 100 gigabytes of data during the Colonial Pipeline ransomware attack.

**Potential Attack Path #3:** Like the first two attack paths, T1059 (Command and Scripting Interpreter), T1557 (Adversary-in-the-Middle), T1040 (Network Sniffing), T1548 (Abuse Elevation Control Mechanism), T1020 (Automated Exfiltration), and T1565 (Data Manipulation) are all relevant for this attack path as well. T1552 (Unsecured Credentials) is not relevant for step 1 of this attack path, as explained for the first attack path. On the other hand, T1087 (Account Discovery) is relevant to discovering credentials for accessing Historian1. Similarly, T486 (Data Encrypted for Impact) is directly relevant since attackers encrypted data to demand ransom during the Colonial Pipeline incident.

Based on this discussion, we observe that the TraCR-TMF was able to identify potential attack techniques from the MITRE ATT&CK matrix and devise attack paths based on them, which could have led to the compromised Business Servers during the actual Colonial Pipeline incident. The cybersecurity experts' validation showed us that most of the attack techniques associated with the steps for each predicted attack path presented in Table VII are contextually reasonable, although the supervised LLM used here was not trained for the Colonial Pipeline evaluation case.

However, there is no way of verifying these predicted attack paths and the associated attack techniques against the actual attack path and associated attack techniques utilized by the attackers during their Colonial Pipeline ransomware attack, because such information is not publicly available. Nevertheless, we can assess whether our framework is able to identify some key exploitations of the actual Colonial Pipeline cyberattack incident. Table VIII presents the exploitations we know about the incident with certainty based on publicly available resources, and the relevance of our framework's prediction to those exploitations, for example, data exfiltration and ransomware. We observe from Table VIII that the TraCR-TMF was able to capture the known exploitations of the actual Colonial Pipeline cyberattack incident through its predicted attack paths and associated attack techniques. This demonstrates the efficacy and transferability of our TraCR-TMF for transportation CPS' threat modeling by validating with a real-world cyberattack incident that falls outside the training domain of the framework's constituent models.

TABLE VII
PREDICTED PATHS FOR COLONIAL PIPELINE RANSOMWARE ATTACK

| # | Predicted Attack Path | Executions Steps using ATT&CK Techniques |
|---|---|---|
| 1 | Human User → VPN Server → Business Servers | 1. The attacker compromises the VPN server using stolen credentials (T1552) or remote command execution (T1059).<br>2. Deploys a malicious payload (T1105) to maintain persistence.<br>3. Uses Network Sniffing (T1040) to gather intelligence about internal traffic.<br>4. Moves laterally to Business Servers using Man-in-the-Middle (T1557) or privilege escalation techniques.<br>5. Executes data manipulation (T1565) or firmware corruption (T1495) on Business Servers. |
| 2 | Human User → VPN Server → Domain Controller1 → Business Servers | 1. The attacker gains access to VPN server using stolen credentials (T1552) or by executing remote commands (T1059).<br>2. Moves to Domain Controller1 using Ingress Tool Transfer (T1105) or Man-in-the-Middle (T1557).<br>3. Gains domain admin privileges and escalates access (T1548).<br>4. Moves laterally to Business Server using Unsecured Credentials (T1552) or Network Sniffing (T1040).<br>5. Executes final payload, exfiltrates data (T1020), or corrupts the system (T1495). |
| 3 | Human User → VPN Server → Remote Access/Jump Server → Historian1 → Business Servers | 1. The attacker exploits VPN server using credential theft (T1552) or executes malicious scripts (T1059).<br>2. Moves to Remote Access/Jump Server via Man-in-the-Middle (T1557) or Network Sniffing (T1040).<br>3. Gains access to Historian1 by using credential access (T1087) or executing remote commands (T1059).<br>4. Uses Historian1 to pivot towards Business Servers via privilege escalation (T1548) or lateral movement techniques.<br>5. Data exfiltration (T1020), encryption for ransom (T1486), or system manipulation (T1565). |

TABLE VIII
COMPARISON OF TRACR-TMF'S PREDICTIONS WITH THE COLONIAL PIPELINE CYBERATTACK INCIDENT FACTS

| Confirmed Events Based on Publicly Available Knowledge | Relevance of TraCR-TMF's Predictions |
|---|---|
| The attackers moved laterally within the Colonial Pipeline IT network | All three predicted attack paths show that an attacker could move laterally within the IT network with the help of different relevant attack techniques |
| The attackers exfiltrated valuable data outside the company's network | Predicted attack paths #2 and #3 include data exfiltration potentials using T1020 technique from the MITRE ATT&CK matrix |
| The attackers demanded ransom after encrypting the company's valuable data | Encryption for ransom or impact using the MITRE ATT&CK technique T1486 is included in the predicted attack path #3 |

## VII. TOWARD PRACTICAL DEPLOYMENT OF TRACR-TMF

In this section, we address four key considerations for the practical deployment of our TraCR-TMF. First, we outline strategies to ensure the framework upholds data privacy in real-world applications. Second, we evaluate the framework's robustness to potential textual errors during prompting. Third, we discuss approaches to keep TraCR-TMF relevant by leveraging the most current data and LLMs, and by sharing codes via publicly accessible repositories. Fourth, we describe how the framework can be adapted for threat modeling in other CPS domains.

### *A. DATA PRIVACY*

The experiments in this study relied exclusively on data from open-source or publicly available frameworks and databases; therefore, data privacy was not a concern. However, in practical deployments of TraCR-TMF within organizational or industrial environments, where sensitive or proprietary data may be analyzed, ensuring data privacy becomes essential. To address this, we present two privacy-preserving adaptation strategies for TraCR-TMF in this section.

#### 1) LOCAL EXECUTION

In this study, we utilized two LLMs, among which ModernBERT is open-source and can be run locally, and GPT-4o is commercial and accessed through OpenAI's Application Programming Interface (API). Organizations with data privacy concerns can solely focus on utilizing open-source LLMs (e.g., oLLaMA, Mistral, GPT4All, or ModernBERT) on their internal datasets and run the TraCR-TMF locally, without transmitting data to external servers. This approach eliminates the need to interact with commercial APIs and provides full control over data handling. While it offers enhanced privacy, performance may vary depending on the capabilities of the selected open-source models.

#### 2) DATA ANONYMIZATION

In scenarios where the use of external APIs (e.g., GPT-based services) is necessary, privacy risks can be mitigated through data anonymization [85] or redaction techniques [86] applied to user inputs prior to query submission. These methods remove or obfuscate identifiers (e.g., IP addresses, organization names) while preserving the essential semantic content required for accurate model interpretation.

### *B. TEXTUAL ERRORS WHILE PROMPTING*

Textual errors can arise at various stages of our pipeline, and the TraCR-TMF is designed to handle them appropriately. These errors may occur within the system prompt, the user-generated prompt, or the input data itself. All the required system prompts for the LLM-based threat modeling based on our TraCR-TMF are provided in this study, minimizing the chance of textual errors during prompting. However, textual errors can still occur in the user prompts when human users are preparing the prompts for their specific use cases or input data, such as the basic input shown in Fig. 2.

To evaluate the robustness of TraCR-TMF against user-side textual errors, we conducted a comparative analysis using clean and synthetically perturbed inputs following the zero-shot approach. The zero-shot approach was considered as it relies solely on the LLM without incorporating any additional preprocessing or downstream filtering, making it an ideal baseline for evaluating robustness against erroneous inputs. In this evaluation, the system prompt (i.e., the prompt presented in Fig. 4, excluding the basic input and the examples) was held constant to eliminate the possibility of introducing typos or misinformation in the system instructions, ensuring that only user-side input, i.e., the basic input presented in Fig. 2, is subjected to perturbation.

We used the nlpaug library [87] to generate three different types of textual noise: (i) typographical errors, (ii) grammatical mistakes, and (iii) synonym substitutions. For each of the five random basic inputs considered for this evaluation, we created 30 noisy variants (10 per noise type) and compared them with the results for the clean input, which

TABLE IX
COMPARISON OF F1 SCORES FOR CLEAN AND NOISY INPUTS

| Input # | F1 Score | | % Change in Avg. F1 Score |
|---|---|---|---|
| | Clean Input | Noisy Input | |
| 1 | CI: [0.200, 0.385] Avg: 0.293 | CI: [0.200, 0.400] Avg: 0.300 | 2.4% |
| 2 | CI: [0.222, 0.222] Avg: 0.222 | CI: [0.158, 0.250] Avg: 0.204 | 8.1% |
| 3 | CI: [0.222, 0.250] Avg: 0.236 | CI: [0.174, 0.256] Avg: 0.215 | 8.9% |
| 4 | CI: [0.235, 0.267] Avg: 0.251 | CI: [0.200, 0.267] Avg: 0.234 | 6.8% |
| 5 | CI: [0.190, 0.268] Avg: 0.229 | CI: [0.188, 0.268] Avg: 0.228 | 0.4% |

CI: 95% confidence interval; Avg: Average were obtained by providing the same clean input to the LLM 30 times. Results in Table IX show that the 95% confidence interval (CI) of F1 scores for clean input #1 is given by [0.200, 0.385], and [0.200, 0.400] for the noisy variants. Thus, there is a certain variance in the result even when the input is free from errors and remains the same across multiple runs. The average F1 score change across all five cases ranged from 0.4% to 8.9%, suggesting that minor textual perturbations did not have a substantial effect on model performance in the zero-shot setting. This demonstrates the robustness of the TraCR-TMF against potential textual errors caused by human users, thereby addressing another practical deployment-focused concern.

### C. MAINTAINING FRAMEWORK RELEVANCE THROUGH UP-TO-DATE DATA, LLMS, AND COMMUNITY COLLABORATION

To maintain the continued relevance and accuracy of TraCR-TMF amid evolving threat landscapes, the framework is designed with modularity, adaptability, and support for dynamic updates. All source codes, trained models, datasets, and utilities are available through a public GitHub repository (see [64]), enabling easy integration, customization, and community-driven enhancements.

TraCR-TMF supports multiple LLM-based approaches, each offering different mechanisms for incorporating up-to-date threat intelligence, as explained below.

- **RAG-based approach:** This approach retrieves current MITRE ATT&CK technique descriptions from a vector database built using the latest MITRE ATT&CK matrix. Code for constructing and updating this vector database is provided in [64], allowing users to dynamically reflect newly published techniques and associated mitigations.
- **ICL-based approach**: This approach relies on a few-shot prompting strategy using representative ATT&CK techniques. As long as examples are curated from the most recent MITRE ATT&CK matrix, model predictions remain current without requiring LLM retraining.
- **Supervised fine-tuning approach**: This approach supports incremental retraining to incorporate newly collected data. Organizations can periodically update models using versioned datasets to maintain alignment with emerging threats and system-specific requirements.

To further support long-term deployment, we recommend periodic expert reviews of the model outputs to identify hallucinations, ensure performance, and inform retraining decisions. The framework is LLM-agnostic, allowing users to integrate newer or more specialized models as they become available.

The GitHub repository of TraCR-TMF [64] is also open to community contributions. Users may fork the repository and submit pull requests for new data, models, or tools. The authors will review the submissions to ensure consistency, correctness, and value to the broader community.

### D. ADAPTABILITY OF TRACR-TMF ACROSS DIVERSE CPS DOMAINS

TraCR-TMF is designed with the inherent characteristics of CPS in mind, enabling its applicability across a broad range of CPS domains, including energy, water, and healthcare. These sectors, despite application differences, share the integration of cyber components, such as computing infrastructure, and physical components, such as sensors and actuators, connected via wired or wireless communication networks. This shared CPS foundation supports the generalization and adaptability of the framework.

TraCR-TMF leverages information on attack techniques, detection strategies, and mitigation measures from the MITRE ATT&CK matrix. The availability of three specialized ATT&CK matrices, namely, (i) Enterprise, (ii) Mobile, and (iii) ICS, enables practitioners to tailor the framework to their specific use case and domain. For instance, a user concerned with wireless communication vulnerabilities may incorporate the Mobile matrix, while a user analyzing OT threats in an ICS may rely on the ICS matrix.

The evaluation of the Colonial Pipeline cyberattack incident presented in this study demonstrates how TraCR-TMF applies to a real-world CPS environment. While our threat modeling focused on the enterprise or IT network of the pipeline and consequently utilized the Enterprise matrix from MITRE ATT&CK, the framework could have similarly supported threat modeling of the Pipeline's OT network using the ICS matrix. Thus, the modular architecture of TraCR-TMF allows users to substitute or combine these matrices based on the CPS context, making the framework readily extensible to diverse domains and applications with corresponding attack surfaces and system architectures.

## VIII. CONCLUSIONS

This study presents TraCR-TMF, an LLM-supported framework for threat modeling of transportation CPS, which requires limited cybersecurity expert intervention. TraCR-TMF leverages LLMs, along with open-source tools and databases, to identify vulnerabilities in transportation CPS, corresponding attack techniques, and potential attack paths that could lead to the compromise of critical assets. By mapping attack techniques to specific vulnerabilities within transportation CPS, the framework empowers cybersecurity professionals dealing with transportation CPS to critically evaluate both proactive and reactive defense strategies. In addition, equipped with insights into relevant attack techniques and paths leading to the compromise of critical transportation CPS assets, cybersecurity professionals can proactively tailor their cyberattack countermeasures. Notably, TraCR-TMF draws from the MITRE ATT&CK matrix to identify applicable attack techniques, providing users with immediate access to established detection and mitigation strategies, which is a valuable starting point for implementing cybersecurity measures. Moreover, TraCR-TMF helps prioritize risk mitigations based on known exploited vulnerabilities to facilitate corresponding threat-informed proactive measures.

TraCR-TMF supports three LLM-based alternative strategies. Of these, the RAG-based approach demands the least cybersecurity expert intervention but is also the least effective in identifying relevant attack techniques. In contrast, both the ICL-based and supervised fine-tuning approaches demonstrate improved performance but require varying degrees of input from cybersecurity experts. While the supervised fine-tuning approach yielded the best performance in this study, benefiting from targeted retraining on the attack identification task with expert cybersecurity input, the performance gap between this method and the other two could narrow as LLMs continue to evolve. In particular, the emergence of models with more advanced reasoning capabilities, such as the anticipated advent of artificial general intelligence (AGI), holds promise for enhancing the effectiveness of less cybersecurity expert-intervention-intensive approaches. In future work, we will explore the integration of such advanced LLMs into the TraCR-TMF.

Although the evaluation scenarios in this study focused on transportation CPS and the LLMs were tailored accordingly, the TraCR-TMF is broadly applicable to CPS across other domains. The tools and databases employed in TraCR-TMF are not specific to transportation CPS and can be applied across various domains. Moreover, the three LLM-supported approaches introduced in this study for mapping existing threats to specific adversarial techniques require varying levels of cybersecurity expert intervention, offering flexible adaptation alternatives to suit the needs of different domains.

Although this study used only public data and raised no privacy concerns, future users may apply the framework to sensitive or proprietary information. To support such cases, TraCR-TMF can be deployed locally with open-source LLMs, eliminating the need to transmit data to commercial APIs while keeping it within the user's secure environment. Alternatively, sensitive inputs can be anonymized before querying commercial LLMs to enhance privacy. These privacy-preserving strategies offer deployment flexibility while preserving the framework's practical utility.

## APPENDIX A
## SELECTION OF F1 SCORE-OPTIMAL THRESHOLD FOR A MULTI-LABEL CLASSIFICATION PROBLEM

We present an operating points selection algorithm (i.e., Algorithm 1) tailored for binary classifiers in a multi-label classification problem, considering a multi-fold cross-validation strategy. For each fold and each label, this algorithm derives the receiver operating characteristic (ROC) curve from the validation set probability scores and corresponding ground-truth labels, thereby yielding false positive rates (FPRs), true positive rates (TPRs), and associated classification thresholds. Subsequently, the F1 score is computed for each threshold by applying binarization to the prediction scores and evaluating against the true labels. The operating point achieving the maximum F1 score is identified; in instances of a unique maximum, it is selected outright. For ties, i.e., wherein multiple thresholds yield equivalent maximum F1 scores, the algorithm resolves ambiguity by selecting the operating point corresponding to the highest TPR among candidates, thereby emphasizing sensitivity while preserving an optimal balance between precision and recall. This method yields fold- and label-specific thresholds optimized for validation performance to facilitate subsequent model evaluation processes. Alternative approaches can prioritize either TPR or FPR based on specific use case requirements. In cases where false positives carry high costs, users may select a threshold that minimizes FPR, potentially at the expense of lower TPR, to reduce erroneous positive classifications. Conversely, when false negatives are more costly, users may choose a threshold that maximizes TPR, accepting a higher FPR to capture most true positives.

## ACKNOWLEDGMENT

The authors would like to acknowledge Dr. Bhavani Thuraisingham, Professor of Computer Science, the University of Texas at Dallas, and Dr. Amjad Ali, Professor of Computer Science, Morgan State University, for their input to this article.

**Algorithm 1** Single-Fold F1-Optimal Operating Points Selector

**Require:** $model_outputs.probs$: $N \times M$ matrix of predicted probabilities from the classifier on the validation set, where $N$ is the number of instances and $M$ is the number of labels

**Require:** $true_labels$: $N \times M$ matrix of true binary labels for the validation set

$operating_points \leftarrow []$
**for** $j \leftarrow 0$ **to** LEN($labels$) $- 1$ **do**
  $scores \leftarrow model_outputs.probs[:, j]$ ▷ predicted probabilities for label $j$ across all $N$ instances
  $y_true \leftarrow true_labels[:, j]$ ▷ true binary labels for label $j$ across all $N$ instances
  $(fpr, tpr, thresholds) \leftarrow$ COMPUTE_ROC_CURVE($y_true, scores$) ▷ calling COMPUTE_ROC_CURVE function
  $f1_scores \leftarrow []$
  **for** $threshold$ **in** $thresholds$ **do**
    $y_pred \leftarrow$ BINARY_PREDICTIONS($y_true, scores$) ▷ applying binarization to the prediction scores by a given threshold
    $f1 \leftarrow$ CALC_F1($y_true, y_pred$)
    $f1_scores$.APPEND($f1$)
  **end for**
  $t \leftarrow$ SELECT_THRESHOLD($f1_scores, thresholds, tpr$) ▷ calling SELECT_THRESHOLD function
  $operating_points$.APPEND($t$)
**end for**

**function** COMPUTE_ROC_CURVE($y_true, scores$)
  initialize $fpr_list \leftarrow []$, $tpr_list \leftarrow []$, $threshold_list \leftarrow []$
  $P \leftarrow$ number of positive samples in $y_true$
  $N \leftarrow$ number of negative samples in $y_true$
  $tpr_list$.APPEND(0)
  $fpr_list$,APPEND(0)
  $threshold_list$.APPEND($\infty$)
  $unique_thresholds \leftarrow$ unique values from scores, sorted in descending order
  **for** $threshold$ **in** $unique_thresholds$ **do**
    $predictions \leftarrow (scores \geq threshold)$
    $TP \leftarrow$ SUM($predictions$ where $y_true == 1$)
    $FP \leftarrow$ SUM($predictions$ where $y_true == 0$)
    $tpr \leftarrow TP/P$
    $fpr \leftarrow FP/N$
    $tpr_list$.APPEND($tpr$)
    $fpr_list$.APPEND($fpr$)
    $threshold_list$.APPEND($threshold$)
  **end for**
  **return** $(fpr_list, tpr_list, threshold_list)$
**end function**

**function** SELCT_THRESHOLD($f1_scores, thresholds, tpr$)
  $max_f1 \leftarrow$ MAX($f1_scores$)
  $indices \leftarrow [index$ **for** $(index, value)$ **in** ENUMERATE($f1_scores$) **if** $value = max_f1]$
  **if** LEN($indices$) $= 1$ **then**
    **return** $thresholds[indices[0]]$
  **else**
    $candidate_thresholds \leftarrow [thresholds[idx]$ **for** $idx$ **in** $indices]$
    $max_tpr_idx \leftarrow$ ARGMAX($candidate_tpr$)
    **return** $candidate_thresholds[max_tpr_idx]$
  **end if**
**end function**

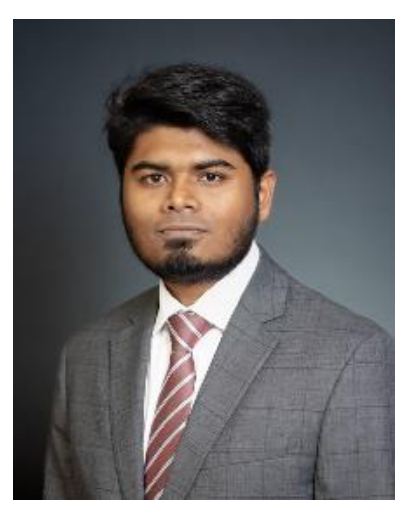

**M Sabbir Salek** (Member, IEEE) received his Ph.D. and M.S. in civil engineering from Clemson University, Clemson, SC, USA, in 2023 and 2021, respectively. Dr. Salek received his bachelor's in mechanical engineering from Bangladesh University of Engineering & Technology (BUET), Dhaka, Bangladesh, in 2016.

Currently, he is working as a Senior Engineer for the USDOT-supported National Center for Transportation Cybersecurity and Resiliency (TraCR) at Greenville, SC, USA. He is also a Senior Research Engineer for the USDOT-supported Center for Regional and Rural Connected Communities (CR2C2), headquartered in Greensboro, NC, USA. He is an Adjunct Faculty with the Glenn Department of Civil Engineering at Clemson University, Clemson, SC, USA. His research interests include cybersecurity and resiliency of intelligent transportation systems (ITS), connected and autonomous vehicle technologies, and advanced, high-performance computing for ITS applications.

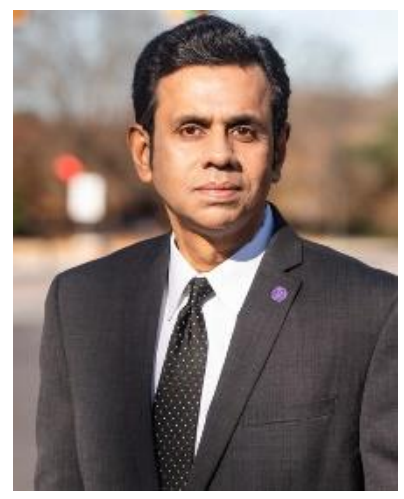

**Mashrur Chowdhury** (Senior Member, IEEE) completed his Ph.D. in civil engineering from the University of Virginia in 1995. Currently, Dr. Chowdhury is serving as the Eugene Douglas Mays Chair of Transportation in the Glenn Department of Civil Engineering at Clemson University. Dr. Chowdhury also holds a joint appointment from the School of Computing (courtesy appointment). Dr. Chowdhury serves as the founding director of the USDOT-sponsored National Center for Transportation Cybersecurity and Resiliency (TraCR) and the USDOT-sponsored Center for Connected Multimodal Mobility ($C^2M^2$). He is also the director of the Complex Systems, Analytics, and Visualization Institute (CSAVI) and a co-associate director of the USDOT-sponsored Center for Regional and Rural Connected Communities ($CR^2C^2$). His current research focuses on the evolving realms of sensing, communications, computing, cybersecurity, and cyber-resiliency, all with the goal of establishing a secure and resilient IoT environment for smart cities and regions.

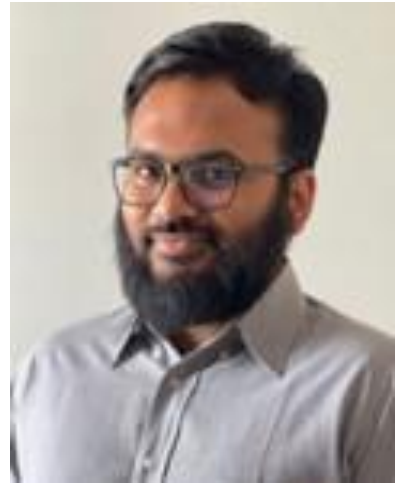

**Muhaimin Bin Munir** (Student Member, IEEE) is doing his PhD in computer science from the University of Texas at Dallas, Richardson, USA. He received his bachelor's in computer science and engineering from Military Institute of Science and Technology, Dhaka, Bangladesh, in 2020.

He is working as a Graduate Research Assistant at the University of Texas at Dallas, where he focuses on cyber threat modeling, multimodal and the application of large language models (LLMs) in cybersecurity. Previously he served as a lecturer at the Military Institute of Science and Technology. His research interests include multimodal LLMs, cyber security and blockchain.

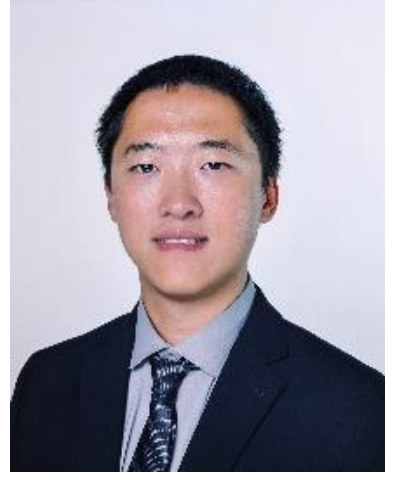

**Yuchen Cai** received his M.S. in computer science from the University of Massachusetts Amherst, MA, USA, in 2020. Yuchen received his B.Mgt. in information management and information system from Beijing University of Technology, Beijing, China, in 2015.

Currently, he is a Ph.D. student in computer science and a Research Assistant at the University of Texas at Dallas, TX, USA. His research interests include Machine Learning, with a focus on Large

Language Models (LLMs), Code Intelligence, and Intelligent Transportation Systems.

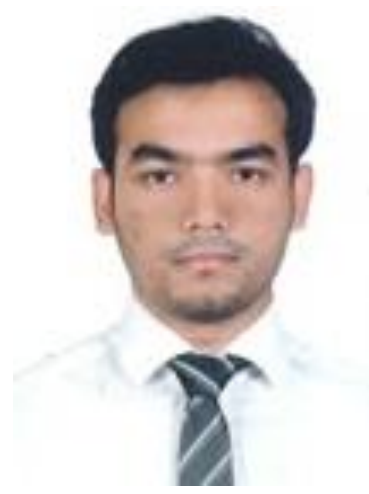

**Mohammad Imtiaz Hasan** (Student Member, IEEE) received his bachelor's in electrical and electronics engineering from Bangladesh University of Engineering & Technology (BUET), Dhaka, Bangladesh, in 2015. Currently, Mohammad is a Ph.D. student in the Glenn Department of Civil Engineering at Clemson University, SC, USA. His research interests include cyber physical system security, quantum computing and Artificial intelligence security.

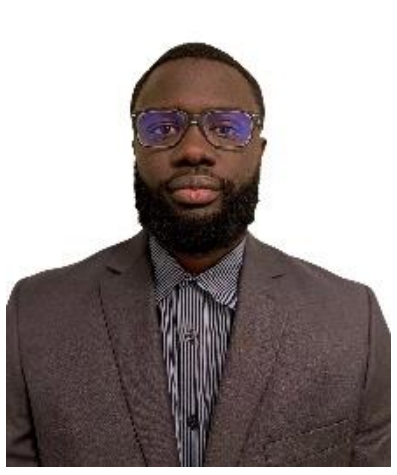

**Jean-Michel Tine** (Student Member, IEEE) is a Ph.D. student at Clemson University, Clemson, South Carolina, USA. He received his M.S. in Civil Engineering from Clemson University from Clemson University in 2025 and a B.S. in Computer Science from Benedict College, Columbia, South Carolina, USA, in 2021.

Currently, he is working as a Cybersecurity Coordinator for the USDOT-supported National Center for Transportation Cybersecurity and Resiliency (TraCR) at Greenville, SC, USA. His research interests include transportation cyber-Physical systems (TCPS), cybersecurity and resiliency of intelligent transportation systems (ITS), connected and autonomous vehicle technologies, and advanced, high-performance computing for ITS applications.

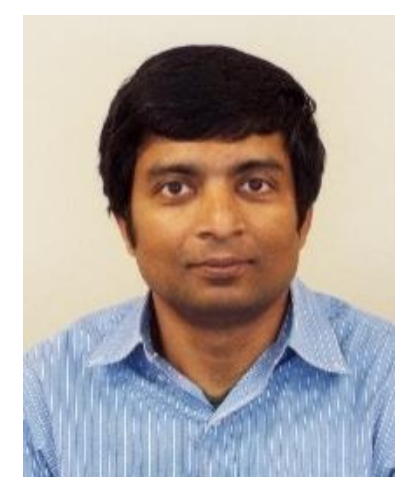

**Latifur Khan** (Fellow, IEEE) is currently a professor (tenured) in the Computer Science department at the University of Texas at Dallas, USA where he has been teaching and conducting research since September 2000. He received his Ph.D. degree in Computer Science from the University of Southern California (USC) in August of 2000.

Dr. Khan is a fellow of IEEE, AAAS, IET, BCS, and an ACM Distinguished Scientist. He has received prestigious awards, including the IEEE Technical Achievement Award for Intelligence and Security Informatics, IEEE Big Data Security Award, and IBM Faculty Award (research) 2016. Dr. Khan has published over 300 papers in premier journals and prestigious conferences. Currently, Dr. Khan's research focuses on big data management and analytics, data mining and its application to cybersecurity, and complex data management, including geospatial data and multimedia data. His research has been supported by grants from NSF, DoT, NIH, the Air Force Office of Scientific Research (AFOSR), DOE, NSA, IBM, and HPE. More details can be found at www.utdallas.edu/~lkhan.

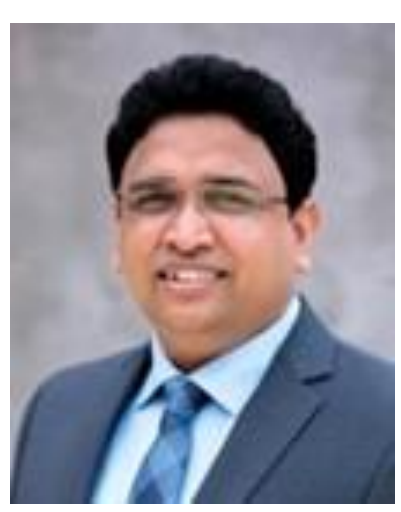

**Mizanur Rahman** (Senior Member, IEEE) received the M.Sc. and Ph.D. degrees in civil engineering (transportation systems) from Clemson University, Clemson, SC, USA, in 2013 and 2018, respectively. He is an assistant professor in the Department of Civil, Construction, and Environmental Engineering at the University of Alabama (UA), Tuscaloosa, AL, USA. His research focuses on cybersecurity and digital twins.